\documentclass[iop,revtex4]{emulateapj}   
\usepackage[titletoc,title]{appendix}
\usepackage{graphicx,natbib,url,twoopt}
\usepackage[varg]{txfonts}
\usepackage{hyperref}               
\usepackage{blkarray}
\usepackage{kbordermatrix}

\renewcommand{\kbldelim}{(}
\renewcommand{\kbrdelim}{)}

\bibpunct{(}{)}{;}{a}{}{,}    

\received{receipt date}
\revised{revision date}
\accepted{acceptance date}
\begin{document}  

\title{pyLIMA : an open source package for microlensing modeling.I. Presentation of the software and analysis on single lens models.}

\author{E. Bachelet$^{1}$, M. Norbury$^{1}$, V. Bozza$^{2,3}$ and  R. Street$^{1}$}

\affil{$^{1}$Las Cumbres Observatory, 6740 Cortona Drive, Suite 102, Goleta, CA 93117 USA}
\affil{$^{2}$\,Dipartimento di Fisica ``E.R. Caianiello'', Universit\`a di Salerno, Via Giovanni Paolo II 132, 84084, Fisciano (SA), Italy}
\affil{$^{3}$\,Istituto Nazionale di Fisica Nucleare, Sezione di Napoli, 80126 Napoli, Italy}

\begin{abstract}
Microlensing is a unique tool, capable of detecting the ''cold'' planets between $\sim$1-10AU from their host
stars, and even unbound ''free-floating'' planets. This regime has been poorly sampled to date owing
to the limitations of alternative planet-finding methods, but a watershed in discoveries is anticipated in
the near future thanks to the planned microlensing surveys of WFIRST-AFTA and Euclid's Extended
Mission. Of the many challenges inherent in these missions, the modeling of microlensing events will be of primary importance, yet is often time consuming, complex and perceived as a daunting barrier to participation in the field. The large scale of future survey data products will require thorough but efficient modeling software, but unlike other areas of exoplanet research, microlensing currently lacks a publicly-available, well-documented package to conduct this type of analysis.
We present first version 1.0 of pyLIMA: Python Lightcurve Identification and Microlensing Analysis\protect\footnote{\url{https://github.com/ebachelet/pyLIMA}}.
This software is written in Python and uses existing packages as much as
possible, to make it widely accessible. In this paper, we describe the overall architecture of the
software and the core modules for modeling single-lens events. To verify the performance of this software, we use it to model both real datasets from events published in the literature and generated test data, produced using pyLIMA's simulation module.
Results demonstrate that pyLIMA is an efficient
tool for microlensing modeling. We will expand pyLIMA to consider more complex phenomena in the following papers.
\end{abstract}

\section{Introduction}     \label{sec:introduction}
At time of writing, 3413 confirmed planets 
have been discovered in various planetary systems. The majority of these planets have been discovered through the transit (2693, with a large proportion thanks to the Kepler/K2 missions \citet{Batalha2013,Coughlin2015}) and the radial velocitiy (609) methods\footnote{\url{http://exoplanetarchive.ipac.caltech.edu/index.html}}.  While other techniques, including astrometry, microlensing and direct imaging, have contributed relatively few detections so far, they probe complementary regions of parameter space.  This large sample of planets has enabled several studies to derive planetary distributions, see for example \citep{Cassan2012, Batalha2013, Clanton2014}, constraining the formation and evolution of planets. The future space missions TESS \citep{Ricker2015} and PLATO \citep{Catala2011} will complete the statistics for this part of the parameter space and will also conduct studies of these new worlds to an unprecedent level of detail, knowing that the atmospheres of these new planets will be perfectly suited for spectroscopic transit follow-up \citep{Ricker2014}. 

However, the transit and radial velocity methods are intrinsically most sensitive to planets orbiting
close their parent star, while direct imaging cannot survey many targets at separations less than
$\sim$10AU, leaving a gap in our understanding of the distribution of "cold" planets at separations between
$\sim$1-10AU where they efficiently form. Fortunately, the sensitivity of the microlensing method peaks around these separations for
galactic host stars and is independent of the lens brightness, meaning that it is uniquely capable of
detecting objects anywhere along the line of sight from Earth to sources in the Galactic Bulge.
Microlensing's capability to complete the planetary census was therefore identified as a priority in last
decadal survey (New Worlds, New Horizons, 2010). In addition, microlensing recently showed its
potential to discover ''free-floating planets'' \citep{Sumi2011}. If these events prove to be caused by truly
unbound planets, then the observed distribution of these objects requires that all planetary systems
eject about two planets, at least one magnitude higher than predictions \citep{Ma2016}.

The WFIRST-AFTA mission has been predicted to detect around 3000 ''cold'' planets by surveying
about 100 million stars in six observing windows of 70 days \citep{Spergel2015}, for a total number of $\sim 37000$ microlensing events. One of the many challenges related to the mission is
therefore the need to model microlensing events in a reasonable time.

However, while the modeling of single lens event is relatively fast and quite easy, the analysis of
multiple lens events – such as planetary or stellar binary systems - is much more difficult. One key
problem is encountered when summing the magnification of the background source images due to $N_l$
multiple bodies in a lensing system, when it is necessary to solve a $N_l^2 + 1$ polynomial for each source
position. Moreover, the lens mapping presents singularities (called caustics) where the magnification of a point source
diverges. The treatment of these singularities, discussed in detail in \citet{Wambsganss1992,Bozza2010,Dong2006}, require methods which are time consuming, and often depend on large
cluster computing facilities.

In addition, the large parameter space that must be searched (at least 7 parameters describe a binary
lens) suffers from several perfect degeneracies, discussed in \citet{Gould2004,Thomas2006,Dominik2009,Skowron2011}, and a number of second order effects (such as parallax and
orbital motion) must be taken into account. Robustly identifying the best-fitting model can be very
challenging, as recent examples can attest (e.g. OGLE-2013-BLG-0723, see \citet{Han2016}).

To date, very few people have developed or used microlensing analysis software. Little systematic
testing has been performed of the existing (proprietary) packages, for which little or no documentation
is available, exacerbating the perceived ''barrier to entry'' for newcomers in the field. This stands in
marked contrast with other areas of exoplanet research, where public codes for detection and analysis
have been developed, and systematic data challenges conducted to stimulate development \citep{Moutou2005,Dumusque2016,Dumusque2017}. This standard of published testing and verification is not only
good scientific practice, it also serves to build confidence in the results.

Our aim is therefore to develop a robust, well-tested and publicly-available software package for the
modeling of microlensing events, capable of answering the needs of future large-scale surveys as well
as existing ground-based programs. Our plan is to devellop the python Lightcurve Identification and Microlensing Analysis (pyLIMA) 
software which adopts the following philosophy :
\begin{itemize}
	\item[$\bullet$]high performance, capable of analyzing multiple-lens events within reasonable time-frames
	\item[$\bullet$]well tested 
	\item[$\bullet$]well documented
	\item[$\bullet$]easy to use
	\item[$\bullet$]open for community use and participation in development
\end{itemize}

The first important step, described in this paper, is to develop a flexible code architecture which
incorporates the fundamental elements of microlensing models and solution finding modules, which
will be built upon as more complex models are introduced in subsequent publications. While the point
source point lens (PSPL) and finite source point lens (FSPL) models presented here are straight-forward, their implementation depends on a number of important elements, including the combination
of data for events observed from multiple facilities and initial-guess assumptions on source stars
colors. These aspects, while rarely discussed, are non trivial, and will form the foundation of the code
for modeling more complex events.

In the  Section~\ref{sec:description}, we outline examples of how such software might be used to derive the design requirements for it, before introducing a general overview of pyLIMA's architecture. The microlensing models are presented in the Section~\ref{sec:models} while the implemented fitting methods are detailled in the Section~\ref{sec:fitmethods}. Section~\ref{sec:realworld} presents the results obtained on both simulated and real datasets. We conclude and present future plans in Section~\ref{sec:conclusions}.

\section{pyLIMA description} \label{sec:description}
\subsection{Use cases examples} \label{sec:user-case}
Here we outline several common scenarios in which the modeling of microlensing events is required, and infer the corresponding requirements placed on the design of pyLIMA.  It should be noted that while only single lenses are considered in the current version of the code, binary and higher-order multiple lenses will be incorporated in subsequent versions, which requires the code design be flexible enough to handle a range of lens types.  

\noindent Use-case 1: ``As a ground-based observer, I have time-series photometric measurements from a number of
different telescopes of an ongoing microlensing event, and I wish to measure the observed and physical
parameters of that event and plot the data overlaid with a model lightcurve, in order to judge whether to continue observations.'' 

This scenario is typical of ground-based microlensing observations where data from multiple longitudinally-separated sites must be combined to fully sample the event lightcurve. We can deduce several requirements from this case:
\begin{itemize}
	\item[$\bullet$]the software should accept mutlitple photometric-timeseries datasets, potentially taken with different filters, to be combined into a single lightcurve, implying the need to properly align the datasets taking into account the different degree of blending (overlapping stellar PSFs or problems with or lack of absolute flux calibrations) in data from instruments with different pixel scales.
	\item[$\bullet$]because each dataset could have its own format, the user needs to be able to describe this format during data input
	\item[$\bullet$]some single-lens models, particularly those including parallax, require information on the location of each observatory, meaning the user needs a way to specify this for each dataset. 
	\item[$\bullet$]the software should output both a text summary of all model parameters and a lightcurve plot with the model overlaid.
\end{itemize} 

\noindent Use-case 2: ``As a new post-graduate, I have access to time-series photometry from a space-based telescope from a
publicly-accessible data archive which includes a number of known single-lens microlensing
events. I have a rudimentary knowledge of both Python, and microlensing theory. I would like to
determine the observed and physical parameters for all of these events, and I have a single desktop
computer with $<$10 CPUs.''

This use-case provides more insight into the circumstances under which the code will be used: 
\begin{itemize}
	\item[$\bullet$]users also have basic but not expert familiarity with microlensing theory, so the documentation should explain the steps of the process and include links to relevant publications. 
	\item[$\bullet$]establishes the use of space-based photometry, telescope resources and survey cadences
	\item[$\bullet$]establishes limits on available computing power -- fitting procedures cannot depend on large-scale parallelization or high-end processors to compensate for efficiency. 
\end{itemize} 

\noindent Use-case 3: ``As the post-doctorate or faculty-level operator of a survey facility producing time-series photometry, I would
like to search for microlensing events within my data, which comprises a database of millions of lightcurves. I have rudimentary knowledge of Python and expert knowledge of microlensing theory, but I do not have the time to fit each lightcurve manually.''

We can outline these needs :
\begin{itemize}
	\item[$\bullet$]establishes different user group which the documentation and user-interface should accomodate.
	\item[$\bullet$]requires that the code be useable as a library, from which the user can choose models and fitting
algorithms to build into software they develop for their own use.
	\item[$\bullet$]implies the fitting procedures must be robust in returning sensible output even for lightcurves which may or may not contain a lensing event.
	\item[$\bullet$]requires that software be able to fit large numbers of lightcurves in an automated manner, which implies that it must be able to robustly establish reasonable initial values for the fitted parameters without human intervention.
\end{itemize} 

\noindent Use-case 4: ``As a professor, my team and I would like to conduct a microlensing survey using telescope facilities available to my institution, which may be both ground- or space-based. I would like to maxmize the science return of this project by simulating the data produced by different possible observing strategies, in order to optimize my use of the facilities.''

In addition to the requirements above, this implies that:
\begin{itemize}
	\item[$\bullet$]the user should be able to specify the characteristics of the telescope(s) and observing strategy to be used, including telescope aperture, location, observing cadence, etc.  
	\item[$\bullet$]the user should be able to specify the range of microlensing parameters for the events to be simulated
	\item[$\bullet$]the software should be able to generate timeseries photometric datasets with realistic noise characteristics and cadence
\end{itemize} 
We note the utility of this simulation module in also providing a means to test the performance of the software itself. 

\subsection{Architecture} \label{sec:architecture}
To be applicable in these use-cases, pyLIMA's architecture needs to be efficient and capable of rapidly analyzing large numbers of events in an automated fashion, but also highly flexible, to enable users to easily conduct detailed analyses.  To address this, pyLIMA's architecture follows three principles:
\begin{itemize}
  \item[$\bullet$] It should be possible to use the software to analyse large datasets in an automatic way. 
  \item[$\bullet$] The code should be constructed in a modular manner, so that users can implement analysis functions of their own design easily simply by adding the desired functions. 
  \item[$\bullet$] The code should be open source and structured in such a way that, as new theories and techniques become available, they can be easily integrated within this framework.  Community contributions are welcomed. 
  \item[$\bullet$] The code should be well documented, to enable new users to learn both microlensing theory and the software functions quickly and easily. 
\end{itemize} 
Python was adopted as the base programming language because it is free, available, and has been widely adopted in astronomy and there are a number of excellent libraries already available (e.g. numpy, scipy and Astropy).  It is also trivial for Python to interface with libraries written in other languages, including C and Fortran. In the following, we  briefly describe the main \textit{modules} that are already implemented in pyLIMA. A more complete description can be found in the pyLIMA documentation\footnote{\url{https://ebachelet.github.io/pyLIMA/}}:

- events : The fundamental starting block of pyLIMA's analysis centers around an Event, which is a \textit{class} with a set of descriptive attributes, including the \textbf{name},  \textbf{ra} and \textbf{dec} (used in the estimation of parallax or extinction along the line of sight).  Since an event may be observed from multiple telescopes and/or with multiple filters, the event can have multiple datasets associated with it.  Each dataset is described as as separate instance of the Telescope \textit{class} (see below).  The principal function of the Event class is \textbf{fit}, which provides the user with a range of options to fit microlensing models to the data. 

- telescopes : this \textit{module} define the \textit{class} Telescope. This \textit{class} groups all the characteristics of an observatory which obtains data on a specific event. The user-specified \textbf{name} and \textbf{filter} distinguish different datasets, representing the filter used for the observations.  Currently pyLIMA supports lightcurve data, which can be provided in units of flux or magnitude, as this \textit{class} provides methods to automatically convert between these units.  The user also specifies a \textbf{location} ('Space' or 'Earth') for the observatory, accepting \textbf{altitude}, \textbf{longitude} and \textbf{latitude} in cases where this is needed for parallax estimation. If the telescope location is 'Space', then the spacecraft position are estimated through the JPL Horizon system : https://ssd.jpl.nasa.gov/horizons.cgi in an automatic manner.

- microlmodels : the user can select a number of different lensing models to fit to the data on a given event, described in the MLModel \textit{class} from this \textit{module}.  The models currently supported are : Paczynski model (point-source point lens model) and finite-source point lens model (FSPL).  Future versions of the software will provide more complex model options.  The user is able to optionally include a range of second order effects, including microlensing parallax, the orbital motion of the lens etc.  The main function of this \textit{class} is to compute the microlensing model associated to the parameters.

- microlfits : A number of well-documented procedures exist for identifying the best-fit model to a given dataset. pyLIMA is structured to provide users with access to commonly-used fitting methods as well as an easy way to implement their own, new techniques if they wish by adding to the MLFits  \textit{class} in the microlfits \textit{module}.  The user is able to indicate their prefered fitting \textbf{method}, which produce outputs appropriate to the \textbf{method} applied. The three \textbf{methods} that are already integrated in the package,  based on the scipy package \citep{Jones2001}, are detailed in the Appendix. Note that several fits, with different models or/and methods, can be performed on the same Event.

- microloutputs : Upon completing an analysis, there are a number of diagnostic plots commonly used in microlensing.  The user is able to produce these using the functions of the microloutputs \textit{module}, which is based on the matplotlib package \citep{Hunter2007}.

\begin{figure*}
  \centering
  \includegraphics[width = 16cm]{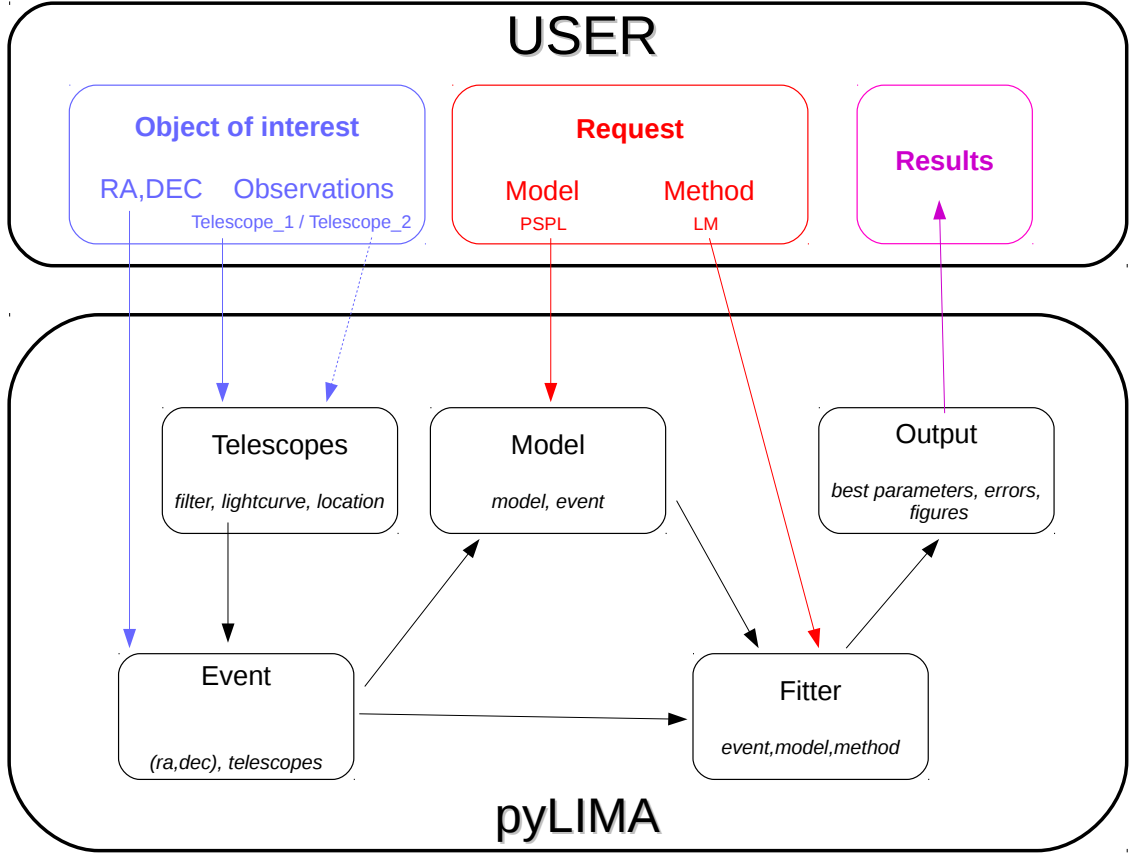}
    \caption{Schematic outline showing a worked example of pyLIMA being used to fit a PSPL model to a typical microlensing event, observed from two telescopes.  In this simple case, the fit is performed using the Leverberg-Marquardt ('LM') method. In blue are the data and informations that the user possesses. The red text indicates the user's commands. The purple box indicates the final products. The black box indicates the different modules used. }
    \label{fig:pyLIMA}
\end{figure*}

- microlsimulator : In addition to fitting real data on microlensing events, a number of important use-cases require the ability to generate simulated data.  pyLIMA provides the microlsimulator \textit{module} for this purpose, incorporating a series of functions which enable the user to produce realistic simulations of how events of given parameters would be observed from the observatories they specify.  This module is based on the astropy package \citep{Astropy2013}.

pyLIMA makes extensive use of the astropy \citep{Astropy2013} and numpy \citep{VdW2011} packages, both of which are being widely adopted in the community, so the functions, attributes and behavior of the software are as familiar as possible.  Figure~\ref{fig:pyLIMA} provides a schematic overview of the architecture, and the code excert below gives an example of these modules in use.  More sophisticated examples are provided in the pyLIMA documentation\footnote{\url{https://github.com/ebachelet/pyLIMA/tree/master/examples}}.

\begin{verbatim}
### First import the required libraries.
import numpy as np
from pyLIMA import event
from pyLIMA import telescopes
from pyLIMA import microlmodels

### Create an event object. 
your_event = event.Event()

### Create two telescopes objects.
data_1 = np.loadtxt('./Survey_1.dat')
telescope_1 = telescopes.Telescope(
	      name='Survey', 
	      camera_filter='I', 
	      light_curve_magnitude=data_1)

data_2 = np.loadtxt('./Followup_1.dat')
telescope_2 = telescopes.Telescope(
	      name='Followup', 
	      camera_filter='I', 
	      light_curve_magnitude=data_2)

### Add the telescopes to your event.
your_event.telescopes.append(telescope_1)
your_event.telescopes.append(telescope_2)

### Construct the model.
model_1 = microlmodels.create_model('PSPL', 
	  your_event)

### Fit using Levenberg-Marquardt algorithm.
your_event.fit(model_1,'LM')

### Producing outputs.
your_event.fits[0].produce_outputs()
\end{verbatim}

\subsection{Good coding practice} \label{sec:coding practice}
When developing software, it is good practice to maintain a clear structure, and to adopt consistent naming conventions for variables, functions etc, to ensure that the code is readable.  This pays off in the long-term in making it substantially easier to maintain and upgrade.  These and other considerations lead to the introduction of the PEP8 standards\footnote{\url{https://www.python.org/dev/peps/pep-0008/}}.  As pyLIMA is intended to be accessible to the entire community, we have adopted this standard for our development, while allowing a certain degree of flexibility.  We have assigned standard astronomical names for variables where they exist, even if they do not respect the PEP8 standards.  An example of this is the $ra$ variable, which in principle should be written as right\_ascension. 

The pyLIMA package is made publicly available via GitHub\footnote{\url{https://github.com/ebachelet/pyLIMA}}, which also facilitates the community contributing to the software development.  Extensive documentation is generated automatically based on mark-up within the code, using \textbf{SPHINX}, making it easier to keep it up to date.  The package also includes some Jupyter notebook-based examples as a guide to users.  

However, as with any substantial package, it is always possible to inadvertantly introduce bugs in the course of implementing new features.  This can be addressed by developing functions which can be run automatically to systematically test all parts of the code and verify that it produces the expected results.  Python provides a framework that enables users to develop these \textit{unit-tests}.  Thanks to the Travis CI portal, these \textit{unit-test} are run automatically through GitHub at each new code upgrade. 
\section{Implementation of PSPL and FSPL models} \label{sec:models}
\subsection{Point Source, Point Lens (PSPL) model} \label{sec:PSPLmodels}
In microlensing, the lens object is in general a star crossing the line of sight between the observer and a background star. The majority of microlensing events detected are in the region of the Galactic Bulge because this densely-populated background field presents the hightest event probability.  Even there, the microlensing optical depth is still low $\tau\sim10^{-6}$ \citep{Udalski1994,Alcock2000,Sumi2011,Sumi2013}.  If the lens is a single massive object, it deflects the light from the background source star into two images separated by several $\theta_E$, the angular Einstein ring radius \citep{Gould2000}:

\begin{equation}
 \theta_E = \sqrt{\kappa M \pi_{rel}} 
\end{equation}

where $M$ is the total lens mass, $\pi_{rel}$ is the lens-source relative parallax and $\kappa$ a constant. For typical events toward the Galactic Bulge, $\theta_E$ is order of few milliarcsec, leading to images that are undistinguishable with current capabilites. However, each image are magnified and then the source flux increase by the total magnification factor $A(t)$, which for a PSPL model is given by \citep{Paczynski1986,Gould2000} :
\begin{equation}
A_{\rm{PSPL}}(t)={{u(t)^2+2}\over{u(t)\sqrt{u(t)^2+4}}} ~ ; ~ u(t)=\sqrt{u_o^2+{{(t-t_o)^2}\over{t_E^2}}}
\end{equation}
where $u(t)$ is the source-lens impact parameter and $u_o=u(t_o)$ is the minimum impact parameter (linked to the maximum amplification $A_o$) at the time $t_o$. $t_E$ is the Einstein ring crossing time :
\begin{equation}
t_E= {{\theta_E}\over{\mu}}
\end{equation}
where $\mu$ is the relative proper motion between the source and lens (i.e. due to the proper motions of Earth, lens and source). It is interesting to note the following properties :
\begin{equation}
A_{\rm{PSPL}}\left\{ \begin{array}{lr} 
\sim{{1}\over{u}} ~; ~u\rightarrow 0 \\
\\
\sim 1 ~;~ u\rightarrow \infty
\end{array}
\right.
\end{equation}
The crowded fields where microlensing is typically observed suffer from a high degree of blending, meaning that the source star point spread function (PSF) usually overlaps those of other stars (including the lens itself).  In combining data provided by telescopes which may have different pixel scales and seeing characteristics, it is necessary to take into account the different degree of blended flux $f_{b,i}$ observed by telescope $i$ as well as the flux from the source, $f_{s,i}$, to give the total flux as a function of time, $f_i(t)$, \citep{Gaudi2012}:
\begin{equation}
f_i(t)=f_{s,i}A(t) + f_{b,i} = f_{s,i}(A(t)+g_i) ~ ; ~ g_i = {{f_{b,i}}\over{f_{s,i}}}.
\end{equation}
Therefore, a PSPL model is described by $3+2n_i$ parameters where $n_i$ is the number of observatories. 

\subsection{Finite Source, Point Lens (FSPL) model} \label{sec:FSPLmodels}
The PSPL models assume that the source is a point. However, this hypothesis breaks down when the source-lens separation becomes small enough to be comparable to the normalised angular source radius $\rho = {{\theta_*}\over{\theta_E}}$ \citep{Yoo2004}. For a typical source star in the Galactic Bulge and a typical lens in the Galactic Disk, $\rho\sim10^{-3}$. This indicates that the effects of finite source size appear when an event 
becomes highly magnified (i.e $u$ approaches zero). Following \citet{WM1994,Yoo2004,Cassan2006}, we used the high-magnification approximation to express the magnification of an extended source with a linear limb-darkening law \citep{Milne1921,An2002} for the wavelenght $\lambda$:
\begin{equation}
A_{\rm{FSPL}}(t,\lambda)=A_{\rm{PSPL}}(t)[B_{0}(z)-\Gamma_{\lambda} B_{1}(z)] ~ ; ~ z = {{u}\over{\rho}}
\label{eq:FSPLmodel}
\end{equation}
where $\Gamma_{\lambda} = {{2u_{\lambda}}\over{3-u_{\lambda}}}$ is the microlensing linear limb-darkening coefficient ($u_{\lambda}$ is the Milne
linear limb-darkening coefficient for the wavelength $\lambda$), $B_{0}(z)$ and 
$B_{1}(z)$ are completely defined in \citet{Yoo2004} and \citet{Cassan2006}.

We implemented a numerical table for $B_{0}(z)$ and $B_{1}(z)$ and their derivative terms within pyLIMA and use a linear interpolation to derive appropriate appropriate coefficients for each computation. The functions $B_{0}(z)$ and $B_{1}(z)$ are define by incomplete elliptic integrals, which are slow and computational intensive to calculate on the fly, explaining our choice of a pre-generated table. This infrastructure makes it straight forward to implement a higher-order limb darkening law if needed in the future. As noted by \citet{Lee2009}, this approach breaks when $\rho\ge0.1$, because the approximation $A\sim{{1}\over{u}}$ breaks for $u\ge0.1$. Fortunately, it is extremely rare to observe $\rho>0.05$ for events in the Galactic Bulge.  \citet{Lee2009} proposed a more robust algorithm but this is more time consuming for most purposes since it requires the computation of double integrals.  Nevertheless, we plan make this algorithm available as an option to the user in future versions.  Currently, the user can either specify values of $\Gamma_\lambda$ for each telescope manually, or allow the code to calculate it automatically from the user-defined telescope filter, source star effective temperature $T_{\rm{eff}}$ and surface gravity $\log g$, using \citet{Claret2011}.

\section{Fitting algorithms} \label{sec:fitmethods}

To date, we have implemented three main solution-finding techniques to fit events.
Note that all methods are applicable to all type of models. 
\subsection{Implementation of Levenberg-Marquardt algorithm}
This method, called 'LM' in pyLIMA, is based on the Levenberg-Marquardt algorithm \citep{Levenberg1944, Marquardt1963}. In python, this function is part of the scipy.optimize.leastsq package which itself is a wrapper  of the C library MINPACK. Following Newton's method, this algorithm uses the gradient to reach the local minimum. The objective function for this method is simply the $\chi^2$ :
\begin{equation}
\chi^2=\sum\limits_i{{(d_i-m_i)^2}\over{\sigma_i^2}}
\end{equation}
where $d_i$ is the i-th data point, $m_i$ is the predicted model value and $\sigma_i$ is the uncertainty on $d_i$. For PSPL and FSPL, we decided to pass the analytical Jacobian matrix (i.e the analytical derivatives of each parameters) to speed-up the convergence. 

In theory the method should stop when the gradient derivatives reach zero, that is, when the algorithm reaches a \textit{critical point}.  However the convergence of this algorithm is set differently. The algorithm stops if : $f_{tol}$ (the relative objective function improvement), $x_{tol}$ (the parameters absolute difference) and/or $g_{tol}$ (the angle between the Jacobian vectors and residuals) are below given thresholds. Note that all of these parameters can be modified easily by the user if needed.

The main difficulty for this method is the requirement for a good initial guess of the parameters. In pyLIMA, the user can provide these guess, but we also developed a method for an automatic estimation. While this can be straightforward for our simulated data, where the noise was 
often negligible relative the signal, it can be difficult for real data. Here we describe the methods we use to estimate each parameter.
\begin{itemize}
  \item[$\bullet$]$t_o$ : This parameter may appear to be the simplest one to find, because a good intial guess should be the brightest point in the lightcurve. However, this approach can easily fail for noisy datasets, so we choose a different approach. Note that we perform this method for all available lightcurves present in the event. For each lightcurve, we first temporarily remove data points with high photometric errors(i.e $\sigma\le \min(0.1,\bar{\sigma}$). Next, we construct a smooth lightcurve using a Savitzky-Golay filter of degree one on the photometry. Then, we perform a loop by selecting points with a flux higher than the actual median and presenting the best photometry as previously described. The loop stops if there is less than 100 points or if the standard deviation of the time is less than five days. $t_o$ for each lightcurve is then the median in time of the remaining points. The final $t_o$ value is chosen to be the mean, weighted by the magnitude uncertainty, for all the telescope's datasets. In fact, weighting the different $t_o$ estimation with the magnitude uncertainty gives more weight to datasets with better photometric precsion, where one would expect this algorithm to have greater success.  
 \item[$\bullet$]$f_s$ : Secondly, we try to find the baseline flux for the survey telescope. This is done in a loop by selecting points below (or within the errobar) of the median flux. Then, by assuming no blending, the source flux for the survey dataset is just equal to the baseline flux. 
 \item[$\bullet$]$u_o$ : Knowing $f_s$, it is possible to compute $A_{max}=\max(flux)/f_s$, the maximum magnification in the survey dataset at the point selected by the $t_o$ estimation, and then $u_o = \sqrt{-2+2\sqrt{1-1/(1-A_{max}^2)}}$.
 \item[$\bullet$]$t_E$ : The estimation of $t_E$ is made using three different methods, the final value being the median of the three. The first method use the fact than when $A=A_{max}/2$, then $t_E=\pm{{t_{demi}-t_o}\over{\sqrt{-2+2\sqrt{1+1/(B^2-1)}-u_o^2}}}$, with $B = {{1}\over{2(A_{max}+1)}}$. The second uses the fact that $A(t_E) = {{u_o^2+3}\over{(u_o^2+1)\sqrt{u_o^2+5}}}$. The algorithm tries to find points closest to this value. The last estimation is a very rough approximation that finds the closest point after/before $t_o$ consistant with the baseline flux and use this as a $t_E$ approximation.
\item[$\bullet$]$f_s ~and~ g$ : These parameters are found for each telescope by using the parameters estimated above, and conducting a linear fit. Note that if $f_s<0$, then the minimum flux value is return as $f_s$ and $g$ are set to zero.
\end{itemize}

\subsection{The differential evolution}
This method is a global optimizer originally presented by \citet{Storn1997}. This method is really robust for a vast range of problems, see for example \citet{Vesterstrom2004}, but is more time consuming than `LM'. This method is called 'DE' within pyLIMA. Again, we used the $\chi^2$ as the objective function. The covergence condition is set by the parameter $tol={{\bar\chi^2}\over{\sigma}}tol>1$ where $\bar\chi^2$ is the mean of the objective function for all population members and $\sigma$ is its standard deviation. We set this parameter to $10^{-5}$. This method is used to find robust guess, which are then used in the method 'LM' presented previously. Note that we slightly changed our fitting strategy in this method. $f_s$ and $g$ are not considered as standard parameters (i.e a full differential evolution search), but they are computed for each step as a linear regression of the flux over the magnification. This is due to the fact that if these parameters are set free, the parameter space volume increases dramatically for a small range of potential correct values. 
\subsection{A Monte-Carlo Markov Chain algorithm}
It is useful to generate the posterior distributions for an event and thus be aware of all plausible models.  We implemented MCMC based on the python module emcee \citep{Foreman2013}. This method requires good initial parameters, which are produced by the method 'DE' in pyLIMA. The user can select whether $f_s$ and $g$ are used as MCMC parameters or computed using a linear regression of the telescope's flux versus the magnification. This method can be called using the 'MCMC' option. Note that this method tries to maximize the log-likelihood, so we set the objective function in this case to be :
\begin{equation}
\log L=-\chi^2/2
\end{equation}

\subsection{Speed performace} \label{sec:speed}
As can be seen from Table~\ref{tab:efficiency}, the median time for the Ground, Space and FSPL fits is of order $0.01$, $0.02$ and $0.1$ seconds respectively. The fits to Space datasets are slower owing to them having higher numbers of points on average. The FSPL dataset is ten times slower for two main reasons. First, the magnification computation is slower due to the linear interpolation necessary to take account of finite source effects. Second, there are two datasets leading to a more complex model and also more data points. Note that the 'DE' method converges in general in about $15$ seconds, which is reasonable given the volume of parameters space it explores. For reference, all our modeling was conducted using an Intel® Core i7 CPU 860 @ 2.80GHz, 8 processor and 16 GB of memory.

\section{pyLIMA fits on various datasets} \label{sec:realworld}
It is crucial to understand how a modeling code performs, we then decide to conduct various tests on both simulated and real datasets. This is a standard method used to test modeling codes on planet detections using the radial velocity and/or the transit method \citep{Diaz2014,Borsato2014,Dumusque2016,Dumusque2017}.The details of simulations can be found in the Appendix~\ref{sec:simulations}, where we also define the quality fits metric.

\subsection{The Ground and Space datasets} \label{sec:ResultsPSPL}
\begin{figure*}   
  \centering
  \includegraphics[width=18cm]{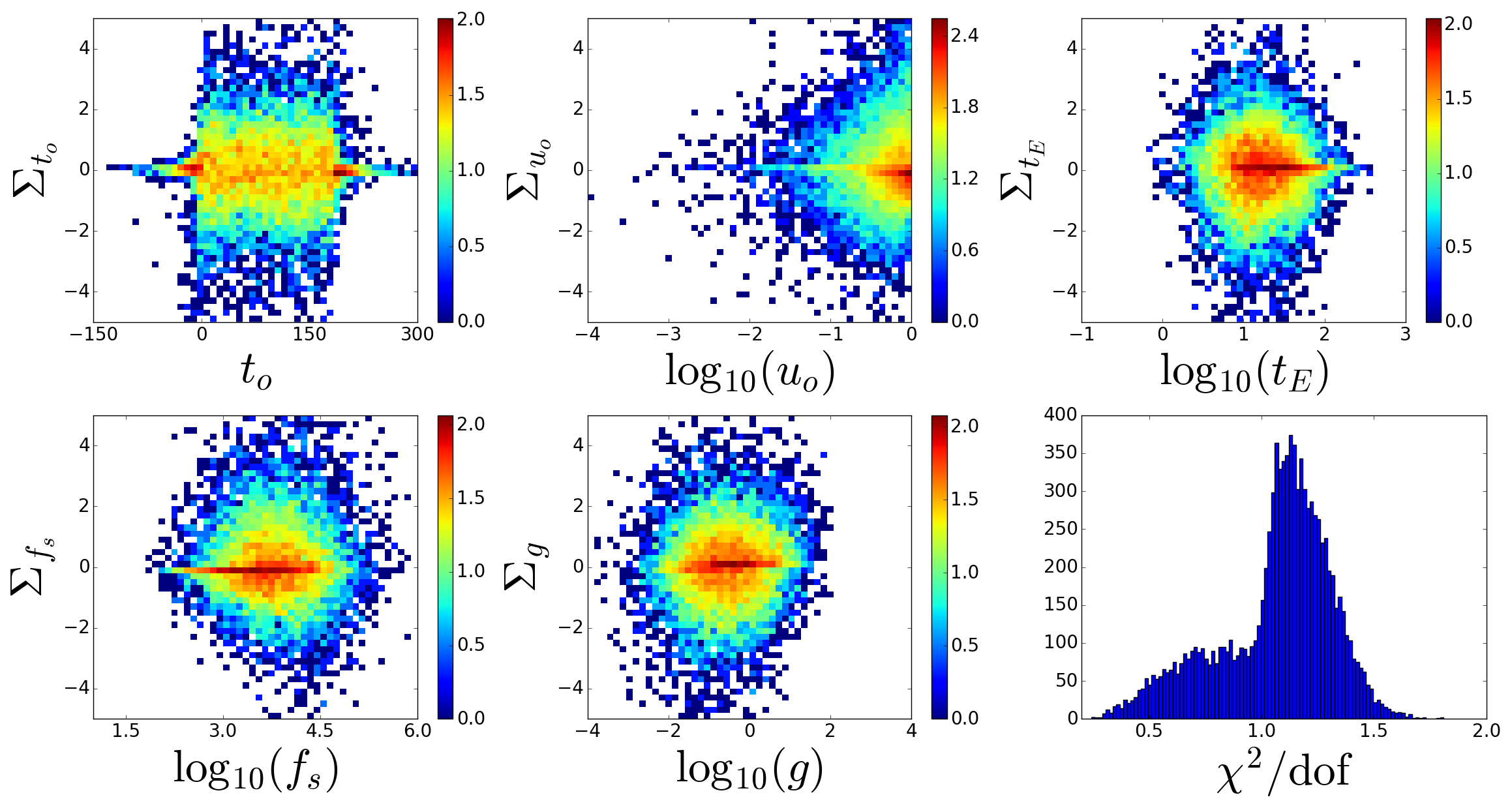}
    \caption{$\Sigma_x$ distributions for the Ground dataset. The x-axis represents the model parameters $x$ and the y-axis represents $\Sigma_x$ limited to -5 to +5. The color scales indicate the $\log_{10}(N)$, where $N$ is the total number of events in the corresponding bin. The bottom right plot is the $\chi^2/ \textrm{dof}$ distribution. A high fraction of fits are consistant with the injected models (i.e $|\Sigma_x|<3$). However, several trends can be observed, especially in the $t_o$, $u_o$ and $\chi^2 / \textrm{dof}$ distributions.}
    \label{fig:Sigmaground}
\end{figure*}

The $\Sigma_x$ results for the Ground and Space datasets can be seen in Table~\ref{tab:efficiency}, Figure~\ref{fig:Sigmaground} and Figure~\ref{fig:Sigmaspace}. From these metrics, it can be seen that pyLIMA accurately recovered the injected models. However, several trends were observed. First, it appears that the fits to the Space dataset are more accurate though at the expense of requiring extra computation time (pyLIMA is about 1.8 slower for this dataset). Both features are explained by the fact that the Space dataset contains more data points (i.e continuous coverage), with a better photometric quality on average (no red noise and photometric precision limited to 0.1\%). Note that this last point also explains why the ${{\chi^2}\over{\rm{dof}}}$ distribution of the Ground dataset is skewed, while the Space one is not. The photometric limitation of 1\% for the Ground dataset induce events with error bar overestimation. This straight leads to a $\chi^2$ understimation as can be seen on the last plot of Figure~\ref{fig:Sigmaground}. Secondly, it may seem strange that $\Sigma_{t_o}$ is minimal when $t_o$ occur outside the observing windows on both datasets, where one would expected that the fitted model should be worse. In absolute terms, the fitted model's $t_o$ estimates are poor but due to the high uncertainty $\sigma_{t_o,\rm{pyLIMA}}$ for these fits, $\Sigma_{t_o}$ vanishes. A discussion about parameter's uncertainties can be found on Appendix~\ref{sec:errorsestimate}. Another interesting trend can be seen in the $\Sigma_{u_o}$ distributions. For the largest model's $u_o$ values, which corresponds to lightcurves with the lower signal, pyLIMA tends to overestimate $u_o$ (i.e underestimate $A_o$, the maximum magnification). This is due to the fact that this kind of lightcurve can be equally well fitted with and without blending flux, as explained in \citet{Thomas2006}.
Note however that this trend is not critical because the fitted values are still less than $1~\sigma$ away from the models.

\begin{table*}
  \scriptsize\setlength{\tabcolsep}{2.5pt}
  \setlength{\columnsep}{1pt}
  \centering
  \begin{tabular}{lcccccccccccccccccccccccccc}
    & \\
     \hline\hline
Datasets&&\multicolumn{3}{c}{Computation time (s)}&&\multicolumn{3}{c}{$\Sigma_{t_o}$}&&\multicolumn{3}{c}{$\Sigma_{u_o}$}&&\multicolumn{3}{c}{$\Sigma_{t_E}$}&&\multicolumn{3}{c}{$\Sigma_{fs}$}&&\multicolumn{3}{c}{$\Sigma_{g}$}&&Method DE\\
    \hline
      & \\
 $\rm{Ground}$&&0.013&0.003&3.306&&62.1\%&86.6\%&94.6\%&&65.5\%&87.5\%&95.1\%&&65.4\%&88.7\%&96.0\%&&66.3\%&87.0\%&94.2\%&&63.1\%&87.6\%&95.6\%&&0\\
 $\rm{Space}$&&0.023&0.005&39.618&&75.1\%&96.2\%&99.6\%&&75.3\%&96.3\%&99.5\%&&74.6\%&96.7\%&99.7\%&&75.8\%&96.0\%&99.3\%&&73.3\%&96.0\%&99.4\%&&1\\
 $\rm{FSPL}$&&0.103&0.019&348.023&&50.3\%&77.7\%&88.6\%&&51.8\%&74.6\%&82.5\%&&50.5\%&77.1\%&88.1\%&&50.4\%&76.5\%&87.4\%&&50.1\%&76.0\%&87.1\%&&853\\
       & \\
    \hline
  \end{tabular}
  \centering
  \caption{pyLIMA fits results for the three simulated datsets. Computation time columns indicated the median, minimum and maximum fit time per lightcurve. Each $\Sigma_x$ column presents the percentage of fits where $|\Sigma_x|$ is less than 1, 2 and 3 respectively. The Space dataset presents the best results, due to the higher number of observations. The FSPL dataset on the contrary presents the worst results. This is due to an intrinsic more complex model than the PSPL model, but also difficulties due to the simulations, see text.}
  \label{tab:efficiency}
\end{table*}

\begin{figure*}    
  \centering
  \includegraphics[width=18cm]{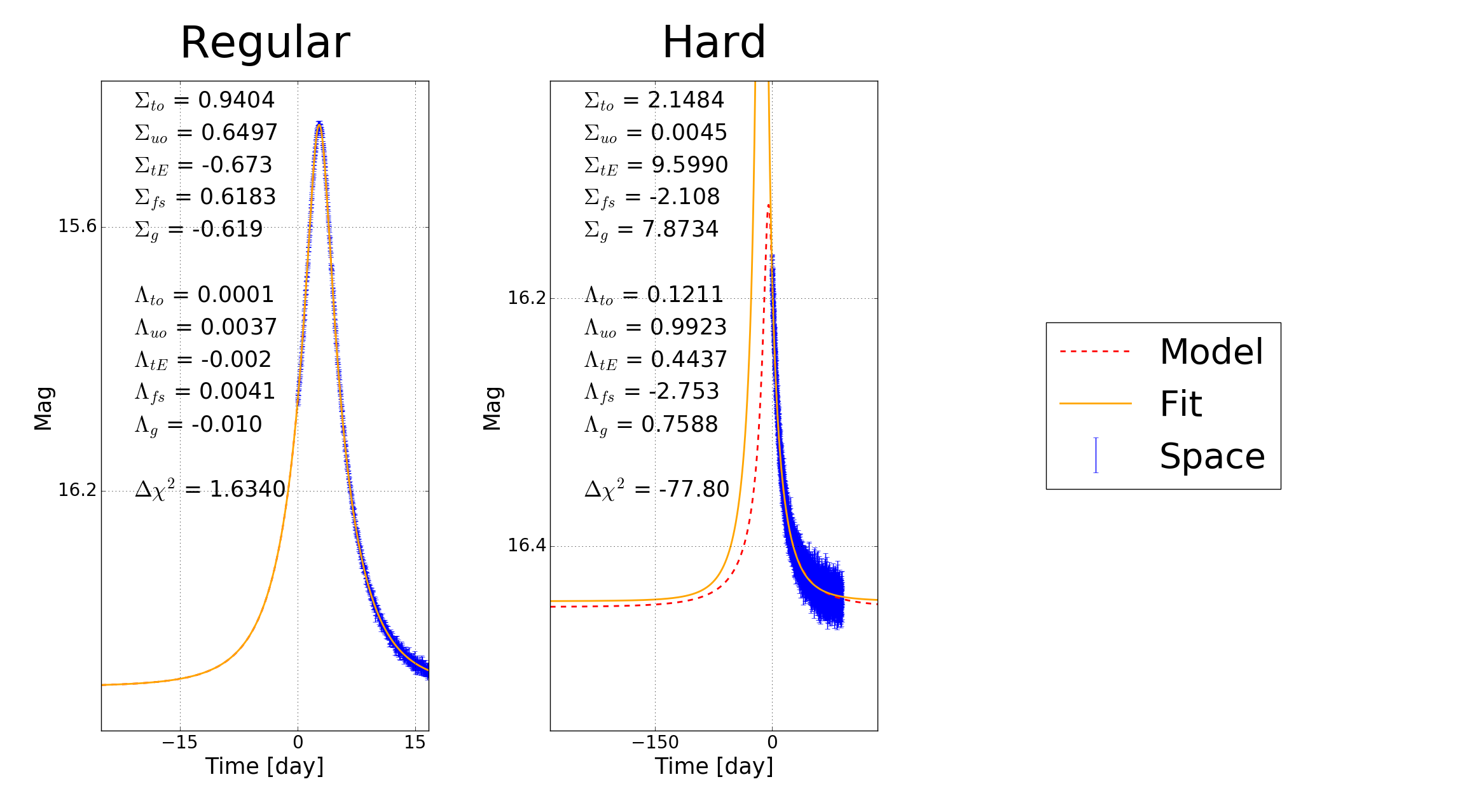}
    \caption{\textit{Left} : An example of 'Regular' lightcurve. The injected model (red) and the fit (orange) are indistinguishable in this case. All metrics indicate a successful fit. \textit{Right} : An example of 'Hard' lightcurve. Various metric indicate that the fit is non-optimal. In this case, the event is not enough constrained by the observations.}
    \label{fig:goodbad}
\end{figure*}

The $\Sigma_x$ shows that pyLIMA accurately fits these datasets in general, but also reveals some unexpected results. In fact, some of the simulated lightcurves were intrinsically very difficult to fit, for example when the observing windows are quite small relative to the event duration. This kind of problem is likely to impact the future space mission. The WFIRST microlensing mission, for instance, will consist of 6 year of ~70 days of observing windows. This is why we limited the observing window to 90 days for the Space datasets. We therefore split the lightcurves into five distinct categories:
\begin{itemize}
  \item[$\bullet$]$Regular$ : The lightcurve doesn't present any of the conditions listed below.
 \item[$\bullet$]$No ~ peak$ : The lightcurve peak occurs outside the observing window.
 \item[$\bullet$]$High ~ blending$ : The lightcurve is highly blended in the model ($g>1$).
 \item[$\bullet$]$No ~ baseline$ : The lightcurve never reaches its baseline during the observing windows ($|t-t_o|<t_E ~ \forall t$).
 \item[$\bullet$]$Hard$ : The lightcurve presents at least two categories listed above.
\end{itemize}
An example of a 'Regular and 'Hard' lightcurves is visible in the Figure~\ref{fig:goodbad} lightcurves for the Space dataset.
\begin{table*}
  \scriptsize
  \centering
  \begin{tabular}{llcccccccc}
    & \\
     \hline\hline
Dataset&Category&$N_{\rm{event}}$&$t_o$&$u_o$&$t_E$&$f_s$&$g$&$\forall x$&$\Delta\chi^2<0$\\
    \hline   
      & \\
&Regular&6017&99.9\%&79.1\%&88.1\%&69.4\%&24.2\%&23.8\%&0.0\%\\
&No peak&1315&66.7\%&32.6\%&69.7\%&33.1\%&4.5\%&3.0\%&0.0\%\\ 
Ground&High blending&2109&98.6\%&49.1\%&59.7\%&43.0\%&36.0\%&35.0\%&0.9\%\\
&No baseline&93&100\%&75.3\%&76.3\%&72.0\%&36.6\%&36.6\%&0.0\%\\
&Hard&466&37.3\%&11.2\%&21.5\%&8.6\%&6.0\%&5.8\%&1.93\%\\ 
       & \\
    \hline
     & \\
&Regular&5180&100\%&93.3\%&96.0\%&86.4\%&44.5\%&44.5\%&0.0\%\\
&No peak&2155&86.4\%&52.5\%&86.4\%&48.8\%&8.0\%&6.4\%&0.09\%\\
Space&High blending&1827&99.8\%&68.1\%&73.7\%&60.8\%&54.6\%&54.5\%&2.57\%\\
&No baseline&104&100\%&69.2\%&69.2\%&69.2\%&31.7\%&31.7\%&0.0\%\\ 
&Hard&734&41.1\%&12.7\%&21.7\%&11.6\%&7.8\%&6.3\%&6.13\%\\ 
       & \\
    \hline
  \end{tabular}
  \caption{pyLIMA's fit success ratio according to the second metric (i.e $|\Lambda_x|<0.1$). Note that the $\forall x$ column indicates the percent of events satisfying $|\Lambda_x|<0.1$ for all parameters. The last column indicates the percentage of fit failures (i.e as a fraction of the total number of events in that category) according to the last metric.}
  \label{tab:efficiency_lambda}
\end{table*}

\begin{figure*}
  \centering
  \includegraphics[width=18cm]{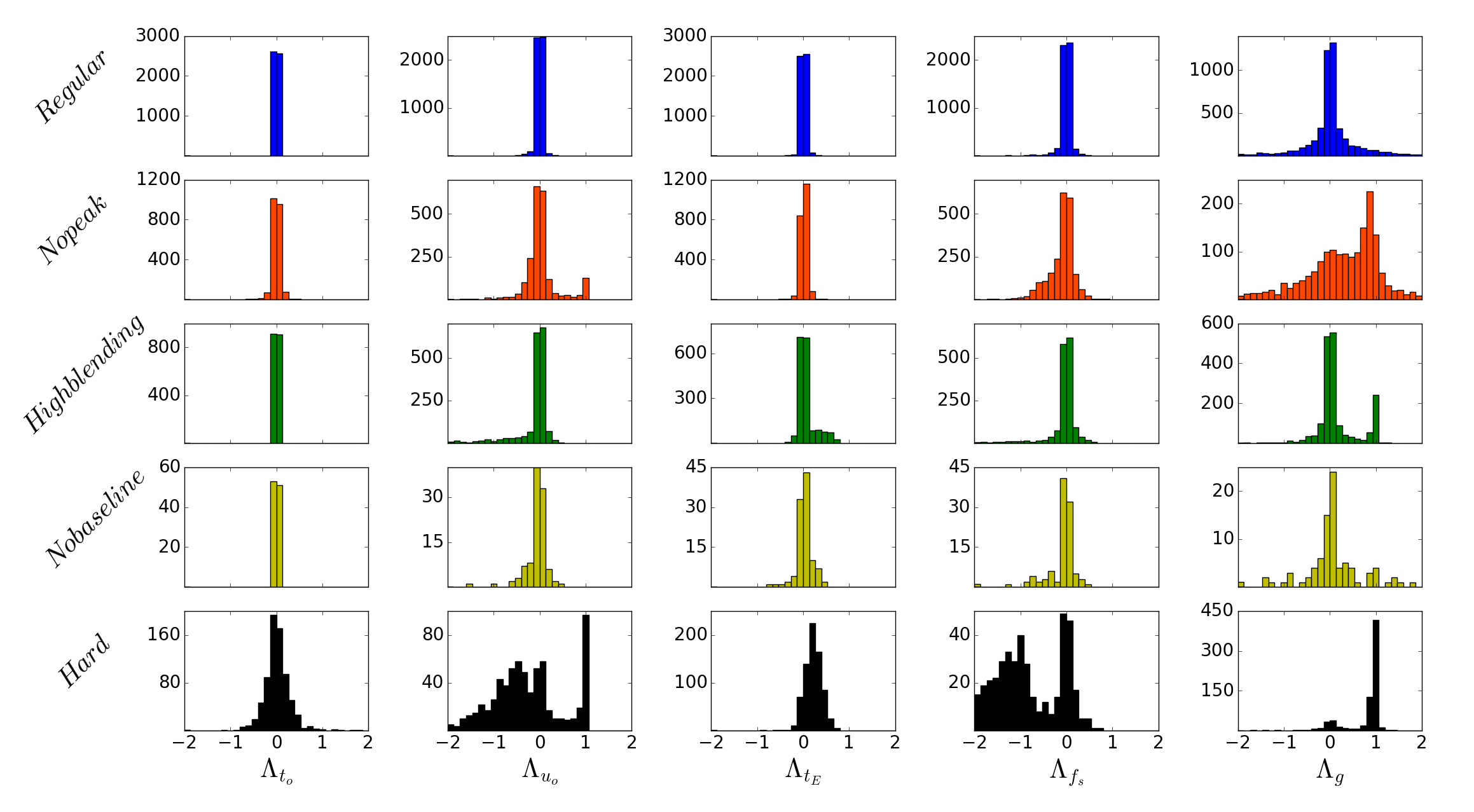}
    \caption{$\Lambda_x$ distributions for the Space dataset. Each category is represented by a color and a row : first row (blue) is \textit{Regular}, second row (orange) is \textit{No peak}, third row (green) is \textit{High blending}, fourth row (yellow) is \textit{No baseline} and last row(black) is \textit{Hard}. It is clear that unsuccessful fits are due to problematic lightcurves rather than pyLIMA fitting routines.}
    \label{fig:lambdaspace}
\end{figure*}
We computed the $\Lambda_x$ metric for each subclass and present our results in the Table~\ref{tab:efficiency_lambda}, Figure~\ref{fig:lambdaspace} and Figure~\ref{fig:lambdaground}. Again, the behaviour of the fits is similar for both datsets and we can characterize the software's performance in each category:

\begin{itemize}

  \item[$\bullet$]$Regular$ : For these lightcurves, pyLIMA accurately recover the model parameters, without any particular trends in the $\Lambda_x$ distributions and without any failures. 

  \item[$\bullet$]$No ~ peak$ : For this subset, it is interesting to note that the parameter $t_o$ is relatively well estimated, without particular trends. However, there are serious trends in the $u_o$, $f_S$ and $g$ distributions. A lot of fits for these events converge to very low $u_o$ values (the orange peak around $\Lambda_{u_o} = 1$ in the Figure~\ref{fig:lambdaspace} and Figure~\ref{fig:lambdaground}). In other words, pyLIMA predicts unrealistically high magnification for these events. These results confirm previous work which indicated that $\chi^2$ minimization algorithms tend to overestimate the magnification if the peak is not observed (see for example \cite{Albrow2004} or \cite{Dominik2009}). This may be related to overfitting. In these cases, the magnification is slightly overestimated, leading to an overstimation of the blending $g$ and so to an underestimation of $f_s$.  This occurs because the lightcurve baseline flux, $f_{\rm{baseline}} = f_s(1+g)$, is in general well constrained (the orange peak around $\Lambda_g\sim1$ and the asymmetry on the left for the orange $\Lambda_{f_s}$ distribution). We found two fit failures for the Space dataset in this category, but considering that $\Delta\chi^2$ equal -0.41 and -0.85 respectively, we do not consider these Failures as critical.

 \item[$\bullet$]$High ~ blending$ : We note that for a large fraction of these lightcurves the fitted model indicated no blending (i.e the green peak around $\Lambda_{g} = 1$). This understimation of the blending is linked to an overestimation of $A(t)$, leading to skewed distributions for $\Lambda_{u_o}$(left) and $\Lambda_{t_E}$ (right). We notice that this category contains a significant number of fit failures for both datasets : 0.9\% (Ground) and 2.57\% (Space). This is due to lightcurves with a very low signal to noise (i.e the lightcurve is nearly flat). These events are particulary difficult to fit, with a median blending values of 7.8 and 5.40 respectively. Note that all these failures come from the method 'LM'.

 \item[$\bullet$]$No ~ baseline$ : It is interesting to notice that this category does not present particular trends. Intuitively, one might expect that the $t_E$ parameter needs baseline observations to be well constrained. As shown by \citet{Dominik2009}, $A(t)\rightarrow{{t_E}\over{|t-t_o|}}$ in the area $u_ot_E<<|t-to|<<t_E$. Therefore, if this region is well sampled, which is the case for this subset, $t_E$ is relatively well constrained.

 \item[$\bullet$]$Hard$ : As expected, this subset presents the worst fitting results. The ratio of failures is the highest for both datasets and fitted parameters are often off from the model (i.e $\Lambda_x\neq 0$). We checked the failed-fit events individually and did not find any critical fit (i.e a fitted model totally in disagreement with the data). The slight disagreement between models and fits (median values of $\Delta\chi^2$ are -5.9 and -3.1 for the Ground and Space dataset respectively) come from an understimation of the blend flux (median values of injected blend flux are 27.4 and 20.6).
  
\end{itemize}

Based on these results, overall pyLIMA's fitting procedure is highly reliable.  The fitted models and estimated uncertainties accurately represent the data for the majority of events. We carefully analyzed the problematic cases and found that instances of poor model fits are a consequence of the intrisic nature of the data (i.e the lightcurve category) rather than a software problem.

\subsection{The FSPL dataset} \label{sec:ResultsFSPL}

\begin{table}
\scriptsize\setlength{\tabcolsep}{2.5pt}
  \centering
  \begin{tabular}{lcccccccccccc}
    & \\
     \hline\hline
Datasets&&\multicolumn{3}{c}{$\Sigma_\rho$}&&\multicolumn{3}{c}{$\Sigma_{fs2}$}&&\multicolumn{3}{c}{$\Sigma_{g2}$}\\
    \hline
     & \\ 
 $\rm{FSPL}$&&52.8\%&74.8\%&82.0\%&&48.0\%&75.2\%&85.7\%&&48.8\%&75.4\%&85.7\%\\
  & \\
    \hline
  \end{tabular}

  \caption{Table~\ref{tab:efficiency} for the three extra FSPL parameterers.}
  \label{tab:efficiencyFSPL}
\end{table}
Table~\ref{tab:efficiency} and Table~\ref{tab:efficiencyFSPL} present the $\Sigma_x$ results for this dataset (distributions can be seen in Figure~\ref{fig:Sigmafspl} and Figure~\ref{fig:lambdafspl}). pyLIMA performed well for this dataset overall, though the fitted model parameters are somewhat less accurately derived than in the previous section. The percentage of events inside the $3~|\Sigma_x|$ windows is lower and pyLIMA required the `DE' method more frequently.   The first trend can be explained by the fact that the fitted parameters errors are $\sim10$ time smaller. For instance, the median values for the  $t_o$, $u_o$ and $t_E$ errors are [$10^{-3},~4\times10^{-4},~9\times10^{-2}$] for the FSPL dataset when they are [$2\times10^{-2},~2\times10^{-2},~4\times10^{-1}$] for the Space dataset. Another explanation for the relative low success of the $\Sigma_{\rho}$ criterion is that for some events, the finite source effect is negligible (i.e when $\rho<u_o$). In these cases, the lightcurves can be equally well fitted with a PSPL model, causing the parameters $u_o$ and $\rho$ to converge far from the model. The second trend is due to the intial guess for $\rho$ produced by the 'LM' method. It is set to $0.05$, which is a very naive and can be far from the correct solution. Note that several alternative initial-guess values were tested, such $\rho={{2}\over{\sqrt{A_{max}^2-1}}}$ \citep{WM1994} or $\rho=u_o$, but this gave similar results due to the rough estimation of $u_o$. It is highly probable that this starting point is too far from the solution, leading to a non satisfactory `LM' fit and causing the software to apply the `DE' method.
\begin{table*}
  \scriptsize
  
  \centering
  \begin{tabular}{llccccccccc}
    & \\
     \hline\hline
Category&$N_{event}$&$t_o$&$u_o$&$t_E$&$\rho$&$f_{s,1}$&$g_1$&$f_{s,2}$&$g_2$&$\forall x$\\
    \hline
      & \\
Success&7497(798)&99.4\%&73.0\%&98.8\%&89.4\%&98.3\%&75.7\%&95.0\%&41.5\%&26.8\%\\
Failures&1650(55)&98.9\%&6.0\%&77.6\%&1.5\%&72.6\%&36.1\%&48.7\%&14.1\%&0.4\%\\
       & \\
    \hline
  \end{tabular}

  \caption{$|\Lambda_x|$ succes ratio for the FSPL dataset. Note that the $\forall x$ column indicates the percent of events satisfying $|\Lambda_x|<0.1$ for all parameters. Numbers in bracket in the first column indicate the number of events where the method 'DE' was use. The low success ratio for $u_o$ and $\rho$ in the 'Failures' category are due to the problematic events already seen in the Figure~\ref{fig:minz} : these events present weak finite-source effects and/or problematic data. Then, it is possible to find competitive models without finite-source effects.}
  \label{tab:efficiency_FSPL}
\end{table*}

\begin{figure}    
  \centering
  \includegraphics[width=8cm]{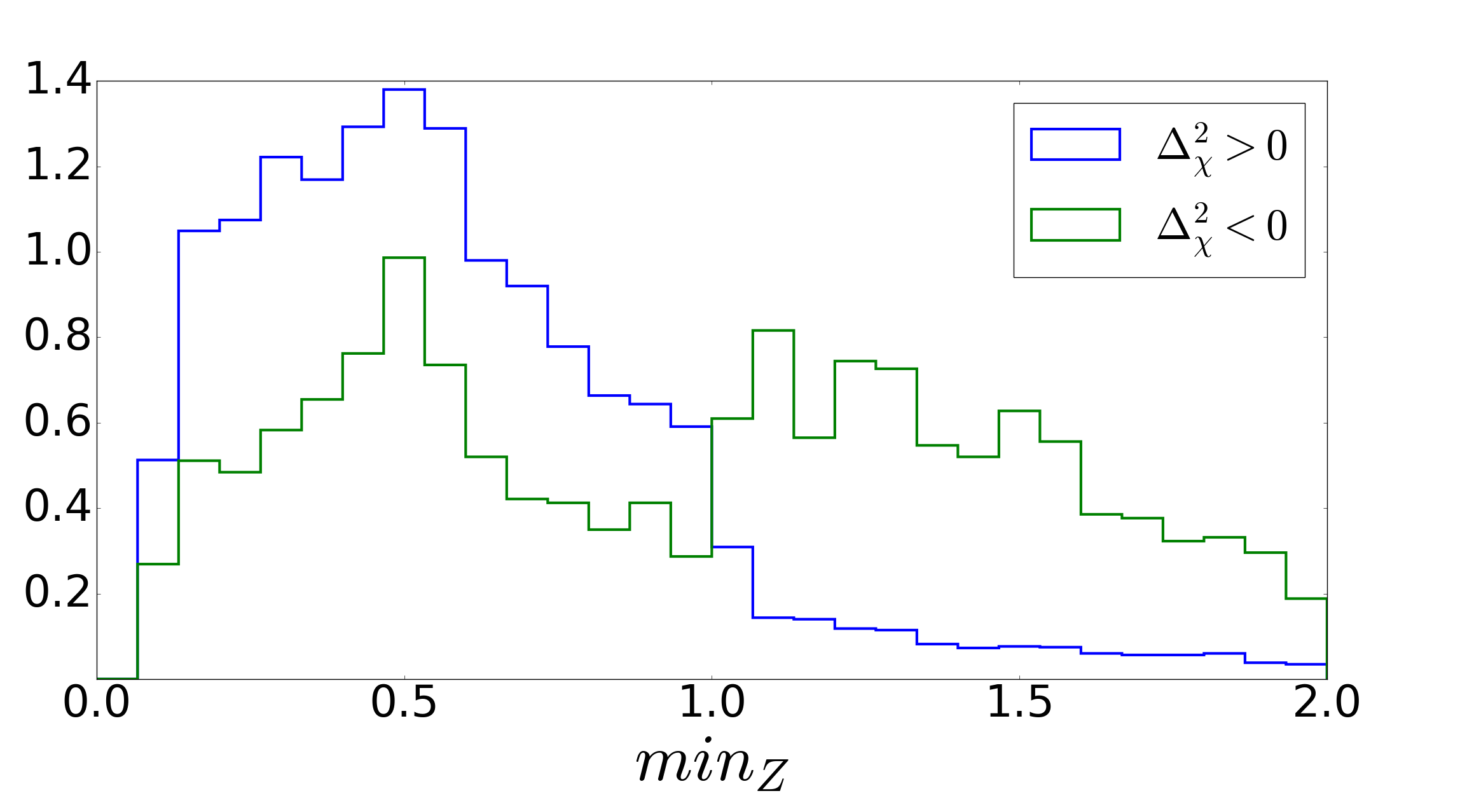}
    \caption{$min_Z$ normalised distributions for successfull fits (blue) and fits failures (green). The significant change for $min_Z\sim1$ indicates that bad fits are more likely to be due to insufficient informations inside the lightcurves instead of fitting routines misbehaviour.}
    \label{fig:minz}
\end{figure}

We report the results for the $\Lambda_X$ metric on the Table~\ref{tab:efficiency_FSPL}. Note that some successful fits have $|\Lambda_{u_o}|>0.1$. This is due to two factors. Some lightcurves present very weak finite-source effects. In this case, the lightcurve can be equally fit with a PSPL model. The second case is linked to the following property of Equation~\ref{eq:FSPLmodel} :
\begin{equation}
A_{\rm{FSPL}}\sim {{2}\over{\rho}} ; z\rightarrow 0
\end{equation}
This means that if $z$ tends to be small, $u_o$ is not constrained whenever $\rho$ is (this explains the difference between the $u_o$ and $\rho$ columns on the first line of the Table~\ref{tab:efficiency_FSPL}). For the fit failures, it is obvious from Table~\ref{tab:efficiency_FSPL} that this comes from an incorrect estimation of $u_o$ and/or $\rho$. However, as we saw previously, some lightcurves could have very weak finite-source effects, making any fit problematic. From Equation~\ref{eq:FSPLmodel}, it is clear that one key fit parameter is $z_o = {{u_o}\over{\rho}}$. In fact, it is more relevant to define $min_z = \min(z)$, the minimum impact parameter sampled by the observations, divided by $\rho$. A distribution of this parameter is presented in Figure~\ref{fig:minz}. We can see that pyLIMA fails when $min_z>1$, in the area where finite-source effects tend to be smaller. It is important to note that more than half of the failures (51.6\%) occurs for an area containing only 17.1\% of the total number of events. For failures occuring for $min_Z<1$, the situation are more complicated, and linked to multiple factors including photometric noise, low sampling, local minimas.   To conclude, it is important to emphasize that the `DE' method is much more reliable (6.5\% of failures) than the 'LM' method (18\% of failures).

\subsection{The OGLE II survey} \label{sec:OGLE2}
To challenge pyLIMA we decided to model microlensing events from the OGLE II \citep{Udalski1994,Udalski2003b,Szymanski2005} survey. We selected events detected in the Galactic Bulge in the three seasons 1998, 1999 and 2000. This leads to 41, 46 and 75 lightcurves respectively. We performed two runs of PSPL modeling, using the 'LM' and 'MCMC' methods.

We discarded lightcurves which present ambiguous behaviour or physical phenomena not yet incorporated into pyLIMA's functionality: binary microlensing, falsely classified as microlensing, variable stars, high photometric noise, etc.  We reject 5, 14 and 10 lightcurves from the respective seasons which leads to a subset of 133 lightcurves in total.  We reviewed the litterature and found several studies which also examined this dataset. \citet{Udalski2000} (U2000) and \citet{Wozniak2001} (W2001) fitted events from the 1998 and 1999 seasons; \citet{Tsapras2003} (T2003) analyzed all three seasons.  \citet{Albrow2000} (A2000) modeled the specific event OGLE-1998-BLG-14 in order to estimate its planet sensitivity.

\begin{figure}    
  \centering
  \includegraphics[width=9cm]{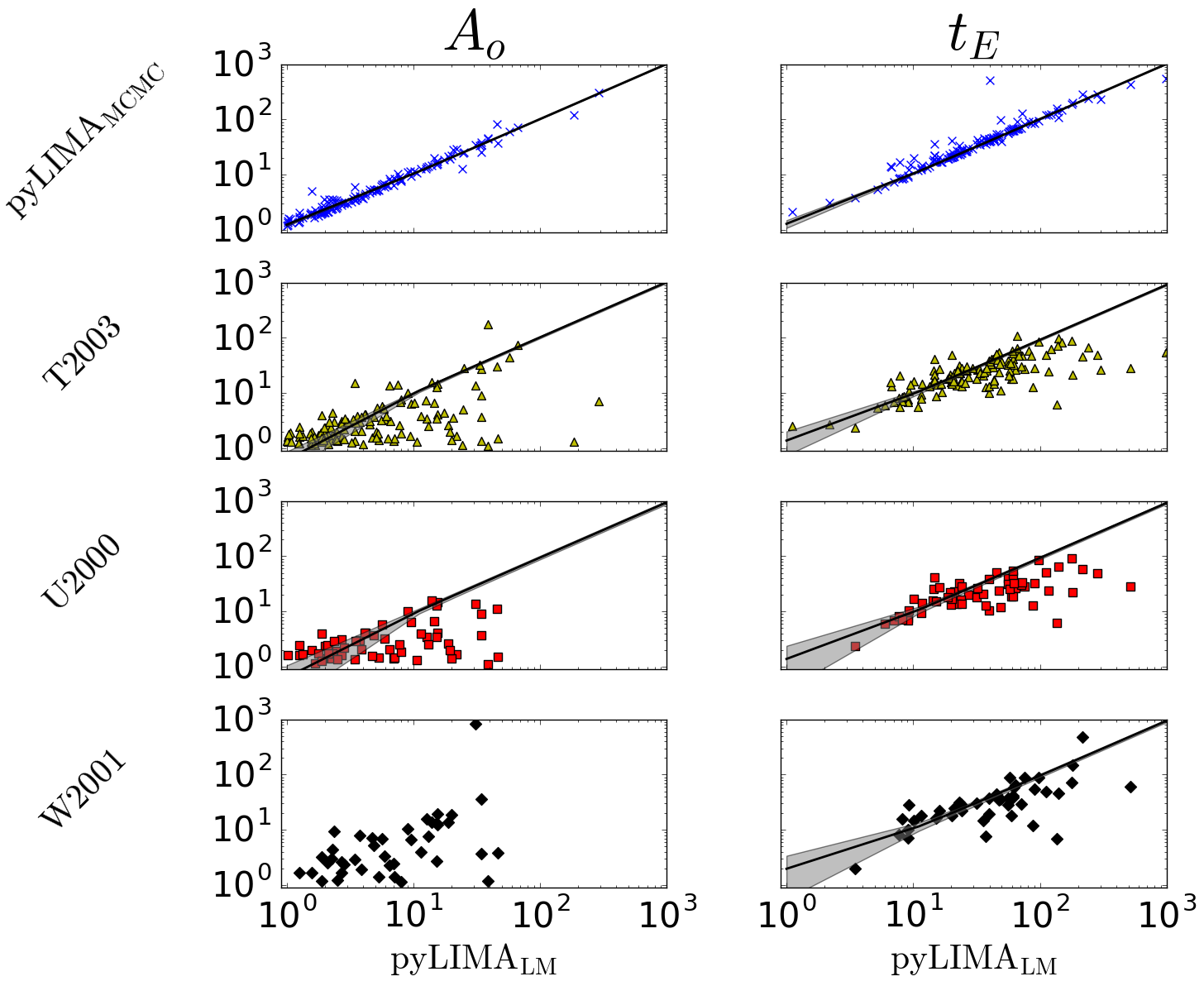}
    \caption{Maximum amplification $A_o$ (left) and Einstein ring crossing time $t_E$ (right) for the OGLE-II survey microlensing events from various studies versus $\rm{pyLIMA_{\rm{LM}}}$. The dark lines are linear fits where coefficient are visible in Table~\ref{tab:PSPL_litterature}. The grey shade indicate the 1 $\sigma$ fit errors.}
    \label{fig:OGLE2comparaison}
\end{figure}

The results of our analysis are plotted in Figure~\ref{fig:OGLE2comparaison} and Table~\ref{tab:PSPL_litterature}. It is interesting to underline that these studies produced various results for $A_o$ and $t_E$, which are the key fit parameters, with the exception of the two methods used by pyLIMA. U2000 and T2003 tend to systematically underestimate $t_E$ and $A_o$ in comparaison with pyLIMA. This is understandable because they do not include blending flux in their fits. W2001 included blend fluxes fitting, leading to better agreement with pyLIMA for $t_E$. For the special case of OGLE-1998-BLG-14, the results between A2000 and pyLIMA 'LM' are in excellent agreement : $\Sigma_{t_o} =-0.10$, $\Sigma_{t_E} =0.10$ and $\Sigma_{A_o} = 0.10$. Note that we use the results of the column 'c' of  the Table 2 in A2000, where the fit was made using OGLE and PLANET follow-up data.

\begin{table*}
  \scriptsize
  \centering
  \begin{tabular}{lcccccccc}
    & \\
     \hline\hline
Study&$N_{\rm{event}}$&$3\Sigma_{t_o}$&\multicolumn{3}{c}{$t_E$}&\multicolumn{3}{c}{$A_o$}\\
 & & & $3\Sigma_{t_E}$ & a & b
 & $3\Sigma_{A_o}$ & a & b \\
 
    \hline
      & \\
$\rm{pyLIMA_{\rm{MCMC}}}$&133&100\%&94.7\%&1.019(0.009)&0.269(0.194)&100\%&1.021(0.010)&0.222(0.055)\\
T2003&132&95.5\%&94.7\%&0.924(0.029)&0.458(0.605)&96.2\%&1.030(0.039)&-0.412(0.210)\\
U2000&52&90.4\%&92.3\%&0.930(0.044)&0.448(0.955)&82.7\%&0.948(0.066)&-0.309(0.346)\\
W2001&43&95.4\%&95.4\%&0.949(0.055)&1.020(1.358)&90.7\%&* & *\\
       & \\
    \hline
  \end{tabular}
  \caption{Comparison between $\rm{pyLIMA_{\rm{LM}}}$ with various studies on the OGLE-II survey. The $3\Sigma_x$ columns indicate the percentage of event where $|\Sigma_x|<3$. The a and b columns indicate the coefficient (and errors) of linear fits (i.e $y=ax+b$) perform on the Figure~\ref{fig:OGLE2comparaison}. A $*$ symbol indicates that the linear fit did not converge. Note the  excellent agreement between the two independent fitting algorithms used in pyLIMA.}
  \label{tab:PSPL_litterature}
\end{table*}
\subsection{Comparison with the ARTEMiS system} \label{sec:artemis}
The ARTEMiS pipeline \citep{Dominik2008} is one of the best real time fitter and anomaly detectors. It is used by the MiNDSTEp group to prioritize their follow-up targets. It performs PSPL fits on all datasets available for each event. We collected all microlensing events from the OGLE 2015 season (and all associated datasets available through the ARTEMiS portal) and compared our fits with those of ARTEMiS. 2145 events were fitted and the results are presented in Table~\ref{tab:ArtemisvspyLIMA}. This time, the $\Sigma_x$ criterion is computed as follow :
\begin{equation}
\Sigma_x = {{x_{\rm{Artemis}}-x_{\rm{pyLIMA}}}\over{\sigma_{x,\rm{pyLIMA}}+\sigma_{x,\rm{Artemis}}}}
\end{equation}
We visually inspected each lightcurve where ARTEMiS and pyLIMA disagreed (i.e $|\Sigma_x| >3$) and found
 four different possibilites :
\begin{itemize}
 \item[$\bullet$]$Anomalous$ : The event is clearly not a single lens microlensing event. It could be due to a binary source, a binary lens microlensing event or a cataclysmic variable for example.
  \item[$\bullet$]$Unknown$ : There is no obvious sign why ARTEMiS and pyLIMA disagree. This could be due to an understimation of errors in both algorithms.
 \item[$\bullet$]$ARTEMiS~ failures$ : The ARTEMiS fit does not agree with the observations.
 \item[$\bullet$]$pyLIMA~ failures$ : The pyLIMA fit does not agree with the observations.
\end{itemize}
We can see that ARTEMiS and pyLIMA agreed about the vast majority of events. The number of failures are low, $1.3\%$ and $1.0\%$ for ARTEMiS and pyLIMA respectively. We note that majority of pyLIMA failures come from a bad estimation of $t_o$. All these failures were made using the method 'LM', and despite the sanity check step, pyLIMA treats these fits as acceptable.
\begin{table*}
 \scriptsize
  \centering
  \begin{tabular}{ccccccc}
    & \\
     \hline\hline
$\Sigma_{t_o}$&$\Sigma_{u_o}$&$\Sigma_{t_E}$&Anomalous&ARTEMiS failures&pyLIMA failures& Unknown \\
    \hline
      & \\
83.7\%,89.8\%,92.3\%&78.5\%,89.4\%,93.1\%&82.1\%,91.4\%,94.5\% & 136& 27& 22& 128 \\
      & \\
    \hline
  \end{tabular}
  \caption{ARTEMiS vs pyLIMA fits results for the OGLE $2015$ season. The percent quantiles are computed on the total of $2145$ events. ARTEMiS and pyLIMA agree on a vast majority of events and only a few fraction (i.e $150/2145$) of events are challenging for pyLIMA, if we consider that all events in the category 'Unknown' come from a bad pyLIMA fit.}
  \label{tab:ArtemisvspyLIMA}
\end{table*}

\subsection{Published FSPL events} \label{sec:realFSPL}
As a final test, we wanted to examine pyLIMA's performance for FSPL events. We found public data for two events : MOA-2007-BLG-400 \citep{Dong2009b} and MOA-2008-BLG-310 \citep{Janczak2010}. These two events present planetary anomalies close to their peaks, however due to the large source sizes relative to the central caustic sizes (i.e ${{\omega}\over{\rho}}<2$), these anomalies have low amplitudes ($\le10\%$) and so an FSPL model is a reasonable fit for almost the entire lightcurves. Finally, we also fit OGLE-2013-BLG-0446 \citep{Bachelet2015} which is a similar event where the authors demonstrated that the hypothetical small planetary signal ($\le1\%$) is probably due to red noise. 

For MOA-2007-BLG-400 and MOA-2008-BLG-310, we used the values of $T_{\rm{eff}}$ given by the authors to compute $\Gamma_I$ and $\Gamma_H$ using \citet{Claret2011}. Note that for these events, the authors used a square-root limb-darkening law, which can explain some fitting differences. For OGLE-2013-BLG-0446, we used $\Gamma_\lambda$ values give in the publication. These results can be found in Table~\ref{tab:realFSPL}.

\begin{table*}
  \scriptsize
  
  \centering
  \begin{tabular}{lcccccc}
    & \\
     \hline\hline
Event&Publication&$t_o-2450000$&$u_o$&$t_E$&$\rho$&$\chi^2(dof)$\\
    \hline
      & \\
&\citet{Dong2009b}$^\dag$&$4354.58107$&$2.5~10^{-4}$&$14.41$&$0.00326$&\\
MOA-2007-BLG-400&$\rm{pyLIMA_{\rm{LM}}}$&$4354.5818\pm{5.5~10^{-5}}$&$3.31~10^{-5}\pm{1.0~10^{-3}}$&$12.98\pm{0.62}$&$0.00363\pm{1.8~10^{-4}}$&1872.49(773)\\
&$\rm{pyLIMA_{\rm{DE}}}$&$4354.5818\pm{5.5~10^{-5}}$&$1.0~10^{-5}\pm{2.3~10^{-3}}$&$14.85\pm{0.78}$&$0.00318\pm{1.81~10^{-4}}$&1854.85(773)\\
& \\
    \hline
& \\
&\citet{Janczak2010}$^\star$&$4656.39975\pm{5.~10^{-5}}$&$3.0~10^{-3}\pm{1.4~10^{-4}}$&$11.14\pm{0.50}$&$0.00493\pm{2.5~10^{-3}}$\\
MOA-2008-BLG-310 &$\rm{pyLIMA_{\rm{LM}}}$&$4656.39904\pm{3.9~10^{-5}}$&$2.84~10^{-3}\pm{1.6~10^{-4}}$&$11.50\pm{0.66}$&$0.00475\pm{2.7~10^{-4}}$&4055.05(3178)\\
&$\rm{pyLIMA_{\rm{DE}}}$&$4656.39904\pm{3.9~10^{-5}}$&$2.84~10^{-3}\pm{1.6~10^{-4}}$&$11.50\pm{0.66}$&$0.00475\pm{2.7~10^{-4}}$&4055.05(3178)\\
       & \\
    \hline

& \\
&\citet{Bachelet2015}$^\spadesuit$&$6446.04790\pm{3~10^{-5}}$&$-4.21~10^{-4}\pm{7~10^{-6}}$&$76.9\pm{1.3}$&$0.000522\pm{~10^{-6}}$&\\
OGLE-2013-BLG-0446&$\rm{pyLIMA_{\rm{LM}}}$&$6446.04660\pm{1.2~10^{-5}}$&$4.0~10^{-4}\pm{3.7~10^{-5}}$&$80.41\pm{7.5}$&$0.000495\pm{4.6~10^{-5}}$&99407.24(2919)\\
&$\rm{pyLIMA_{\rm{DE}}}$&$6446.04660\pm{1.2~10^{-5}}$&$4.0~10^{-4}\pm{3.7~10^{-5}}$&$80.09\pm{7.4}$&$0.000497\pm{4.6~10^{-5}}$&99407.32(2919)\\
       & \\
    \hline
 
  \end{tabular}
   \\
  $^\dag$\scriptsize The parameters are from the binary 'Close' model.\\
  $^\star$\scriptsize The parameters are from the binary 'Wide' model.\\
  $^\spadesuit$\scriptsize The parameters are from the FSPL model in their Table~3.
  \caption{pyLIMA fits results for the three FSPL like events. Results are in good agreement with the literature.}

  \label{tab:realFSPL}
\end{table*}
Again, the results are in good agreement with the literature. The $\chi^2$ of pyLIMA fit for OGLE-2013-BLG-0446 is higher, but these calculations are made without any errorbar rescaling. As shown in \citet{Bachelet2015}, some telescopes need high errorbar rescaling to obtain a normal distribution of the fit residuals.
\section{Conclusions and recommendations} \label{sec:conclusions}
In this paper, we describe the first phase of the development of pyLIMA, an open source microlensing analysis package. 
We present the method and tools we used to build a flexible and easy to use architecture, accessible to all users.  We have conducted a series of test to assess the reliability of pyLIMA fitting two single lens models, both on simulated and real datasets. To do so, we define three different metrics to assess the quality of the fits.

Results on the simulated Point-Souce Point-Lens events are statisfying both in terms of fit convergence ($\ge$ 99\%), parameters and uncertainties estimation. The complete analysis of our simulations reveals several trends already discovered in the litterature \citep{Albrow2004, Thomas2006, Dominik2009}. This study also highlights potential issues of future microlensing space missions due to their relative short observing windows (i.e $\le$ 100 days) every year. We also used pyLIMA to fit the OGLE-II dataset and found a good agreement with four previous studies. Finally, we found an excellent agreement between pyLIMA and the ARTEMiS system on the OGLE-IV 2015 microlensing season (i.e $\ge$ 90 \% match).

We also implemented the Finite-Souce Point-Lens model and runs similar tests. Despite a good agreement, we found a higher failure occurence ($\sim$ 18\%). This is due to two main reasons. The first one is due to the difficulty to find a "good enough" estimation of $\rho$ when using the 'LM' method. The second one is due to some difficult lightcurves where the finite-source effects are low (i.e $z>1$). The failure rate drops significantly when the `DE' method is used (6.4\% versus 18\%).
 
Based on these results, the authors make some recommendations to users. The fitting of lightcurves with PSPL models can be made using the `LM' method with a good expectation of fit convergence. This enables the user to study large datasets in a reasonable ammount of time.  In case of doubts or the failure of a fit, the `DE' method  should be used. Fitting FSPL models needs a bit more caution. If the lightcurve is well sampled and exhibits strong finite-source effects, the `LM' method should converge. In other case, users are encouraged to use the method 'DE'.  Finally, the authors recommand the use of the 'MCMC' method only in two cases. First, the 'MCMC' method is used to derive the event parameters posterior distributions. This give a more complete view to the parameter space than the 'LM' method. Secondly, if the event is poorly constrained, it is likely that the Fisher matrix inversion will return unrealistic error. Then the 'MCMC' approach should return a more comprehensive view of the problem. To conclude, we would like to notice that this have a cost, since the 'MCMC' method is about 2000 slower than the 'LM' (or 'DE') method, since it requires the computations of thousands of models.


\section*{Acknowledgements}
The authors thank the referee for his/her useful comments. The authors express their gratitude to the microlening observer teams, especially to the OGLE Survey, for kindly granting access to their photometry. This research has made use of NASA's Astrophysics Data System. Work by EB and RAS is support by the NASA grant NNX15AC97G.

\appendix


\section{Description of simulations} \label{sec:simulations}


\subsection{Description of the datasets} \label{sec:PSPLsimulations}
\subsubsection{The Ground dataset} \label{sec:Groundsimulations}

The first dataset, called Ground, mimics PSPL observations from a unique terrestrial survey such as OGLE \citep{Udalski2002} or MOA \citep{Bond2001}.
The observing window was arbitrarily set to [0, 182] days for each lightcurve. This corresponds approximately to a one year observation of the Galactic Bulge from the South hemisphere. We first implemented a night/day cycle and we selected the number of exposures per night from a uniform distribution between
$1$ and $30$ to mimic survey observations of different fields with different cadences. To make these simulations more realistic, we implemented potential bad weather. 10 \% of nights were randomly selected to be "bad weather" (no observations). We also implemented a full moon avoidance windows ( $5$ days in a row). However, to ensure that a microlensing event would be detected, we ensured that at least two points were observed around the event peak $t_o$.  We decided to implement Poisson and red noise sources, the latter with a sum of low amplitude ($\le5\%$) and low period ($\le10$ days) sinusoidal functions. The photometric precision was limited to a minimum of $1\%$.  $u_o$ was selected from a uniform distribution between $10^{-4}$ and $1$, while $t_E$ was selected from a log-normal distribution ($\mu=2.8$, $\sigma=0.9$), which is a rough approximation of the expected one (see for example \citet{Sumi2011}). $t_o$ was generated from a uniform distribution between $-t_E$ and $182+t_E$ days and the source and blend magnitudes were produced from a normal distribution of ($\mu=18$, $\sigma=1.5$) and ($\mu=19.4$, $\sigma=1.6$) respectively.
\subsubsection{The Space dataset}\label{sec:Spacesimulations}

The second dataset, called Space, reproduces PSPL observations from a space-based survey such as WFIRST \citep{Spergel2015} or EUCLID \citep{EUCLID2011}. The data were simulated in the same way as above, except that we implemented continuous coverage (no night/day cycle), we restricted the photometric precision to a minimum of $0.1\%$, we set the observing windows to [0, 90] days and we fixed the observation sampling to 30 minutes. This is an approximation of the expected duration of the annual WFIRST Bulge surey. We also removed red noise effects from our simulations. 

\subsubsection{The FSPL dataset}\label{sec:FSPLsimulations} 
For this dataset, we simulated two telescopes (Survey and Follow up) for each FSPL event. This represents the fact that finite source effects are detected only in high magnification events, which are a priority for the follow-up teams such as RoboNet, PLANET, MiNDSTEp or $\mu$FUN because they are highly sensitive to planets \citep{Griest1998}. In these events, the effects of finite angular source size are seen only close the magnification peak and have a short duration (from hours to a few days).  Using two telescopes is also consistent with the design goal that pyLIMA should be generally applicable to all ground-based and space-based datasets. Since until recently a lot of microlensing events have been covered by a combination of surveys and follow-up data, and since this is the most challenging combination of data, this is an excellent test case for the code. The Survey observatory is similar to the Ground simulations presented above. This dataset represents a single site survey, which contains a daily gap in the lightcurve due to the day/night cycle. To ensure that finite source effects were really present in our simulations, the Follow up data consists of two days of observations around $t_o$ in order to catch deviations from a PSPL model. The source magnitude and blending ratio of the Follow up observations are chosen to be different from those of the 'Survey' data in order to reproduce the fact that these follow-up telescopes have different spatial resolutions, sky conditions and photometric reduction pipelines. We also choose a random cadence of observations (between 0 and 30) for the Follow up dataset. We decided to implement Poisson noise and red noise for both telescopes. To ensure that finite source effects were observed, we forced $t_o$ to be inside the observing window [0 to 183] days. We selected $\rho$ from a uniform distribution from ${{1}\over{2}}u_o$ to $\min(10~u_o,0.05)$. We limited the unifom distribution of $u_o$ from $10^{-4}$ to $0.025$ and the photometric precision to $1\%$. For the purposes of this paper, we set $\Gamma_{\lambda} = 0.5$ for the simulation and the fitting process. Again, ten thousand events were simulated.

\subsection{Definition of fit quality metrics} \label{sec:metrics}
To evaluate the accuracy of the fits, we compared the results from pyLIMA to the injected models in the lightcurves. It is true to say that the best fit model is never identical to the injected model due to the noise in the data and the discontinuous sampling. However, it is also true that the input model represents the observations accurately. Therefore, a comparison of the injected versus the best-fit model parameters can be used to test the robustness of the fit (i.e to estimate the accuracy of the parameters derived from the fit). We defined three differents metrics to analyze the fits of our simulated events.

We first defined the $\Sigma_x$ metric :  
\begin{equation}
\Sigma_x = {{x_{\rm{model}}-x_{\rm{pyLIMA}}}\over{\sigma_{x,\rm{pyLIMA}}}}
\end{equation}
where $x$ is the model parameter and $\sigma_{x,\rm{pyLIMA}}$ is the error on the parameter returned by pyLIMA. For the method 'LM', $\sigma_{x,\rm{pyLIMA}}$ is the square root of the parameter's variance obtained from the fit covariance matrix. In future version of this software, this approximation will be replaced by more robust methods to estimate the variance, such as the bootstrap technique. It is also informative to judge predictions made by the fit for the future evolution of an event. For example, if an event has not yet reached its peak, we can judge how well pyLIMA  estimates the microlensing parameters (i.e. we can define a good prediction if $|\Sigma_x| <1 ~;~\forall x$ for example). It also reveals whether parameter errors are correctly estimated. However, this metric suffers one strong flaw : it vanishes when the parameter error diverges, which can happen with the covariance matrix estimation (see Section~\ref{sec:errorsestimate}).

To counter this problem, we defined a second metric $\Lambda_x$ :
\begin{equation}
\Lambda_x = {{x_{\rm{model}}-x_{\rm{pyLIMA}}}\over{\epsilon_x}};  \left\{
\begin{array}{lr}
\epsilon_{t_o} = t_{E,\rm{model}} \\
\epsilon_{u_o} = u_{o,\rm{model}} \\
\epsilon_{t_E} = t_{E,\rm{model}} \\
\epsilon_{f_s} = f_{s,\rm{model}} \\
\epsilon_{g} = g_{\rm{model}} \\
\end{array}
\right.
\end{equation}
This metric is then a relative error and is insensitive to the estimation of parameter uncertainties. We opted to consider a fitted value acceptable if $|\Lambda_{x}| <0.1$.

Finally to judge whether or not a fit is a succes, we computed:
\begin{equation}
\Delta{\chi}^2 = {\chi^2_{\rm{model}}} - {\chi^2_{\rm{pyLIMA}}}
\end{equation}
This is a more robust statistic than ${{\chi^2}\over{\rm{dof}}}$ because the latter could be insensitive to a bad fit where the "area of interest" (i.e the microlensing event) is not significant regarding the total lightcurve length. We finally defined a fit successful if $\Delta{\chi}^2>0$. However, we also computed the ${{\chi^2}\over{\rm{dof}}}$ and present this in the following sections for completeness.

\section{Discussion on the parameters errors estimation} \label{sec:errorsestimate}

In this appendix, we study in more details the behaviour of the parameters error estimations. Error estimations for the methods 'LM' and 'DE' come from the inverse of the Fisher matrix whereas the error estimations for the 'MCMC' method come from the 1 sigma confidence interval around the distribution median. The Fisher matrix can be written as \citet{Tsapras2016}:
\begin{equation}
F_{i,j} = \Bigg \langle \Bigg({{\log{L}}\over{dp_i}}\Bigg)\Bigg({{\log{L}}\over{dp_j}}\Bigg) \Bigg \rangle
\end{equation}
where ${\log{L}}$ is the log-likelihood function. In the case where the noise model is assumed indepedent of the model, uncorrelated and normally distributed \citep{LSST2009}, the log-likelihood can be write as:
\begin{equation}
\log{L} = -{{1}\over{2}}\sum_{n}{\log{2\pi}}-{{1}\over{2}}\sum_{n}{{{(d_n-m_n)^2}\over{\sigma_n^2}}}
\end{equation}
Assuming two parameters $(p_i,p_j)$ of the model $m$, we can rewrite the Fisher matrix:
\begin{equation}
F_{i,j} = \Bigg \langle \sum_{k}{{(d_k-m_k)}\over{\sigma_k^2}}{{dm_k}\over{dp_i}}\sum_{l}{{(d_l-m_l)}\over{\sigma_l^2}}{{dm_l}\over{dp_j}} \Bigg \rangle
\end{equation}
Since $\langle(d_k-m_k)(d_l-m_l) \rangle = \sigma_k^2\delta_{k,l}$ \citep{LSST2009}, with the Kronecker delta function noted $\delta_{k,l}$, the Fisher matrix reduce to \citep{Mogavero2016}:
\begin{equation}
F_{i,j} =\sum_{n}{{1}\over{\sigma_n^2}}{{dm_n}\over{dp_i}}{{dm_n}\over{dp_j}}
\end{equation}

The details for each PSPL parameters are listed below :
\begin{itemize}
\item[$\bullet$] ${{df}\over{dt_o}} = -f_s {{dA}\over{du}}{{t-t_o}\over{t_E^2u}}$
\item[$\bullet$] ${{df}\over{du_o}} = f_s {{dA}\over{du}}{{u_o}\over{u}}$
\item[$\bullet$] ${{df}\over{dt_E}} = -f_s {{dA}\over{du}}{{(t-t_o)^2}\over{t_E^3u}}$
\item[$\bullet$] ${{dA}\over{du}} =  {{-8}\over{u^2(u^2+4)^{3/2}}}$
\item[$\bullet$] ${{df}\over{df_s}} = A+g$ 
\item[$\bullet$] ${{df}\over{dg}} = f_s$
\end{itemize}

\begin{figure*}    
  \centering
  \includegraphics[width=18cm]{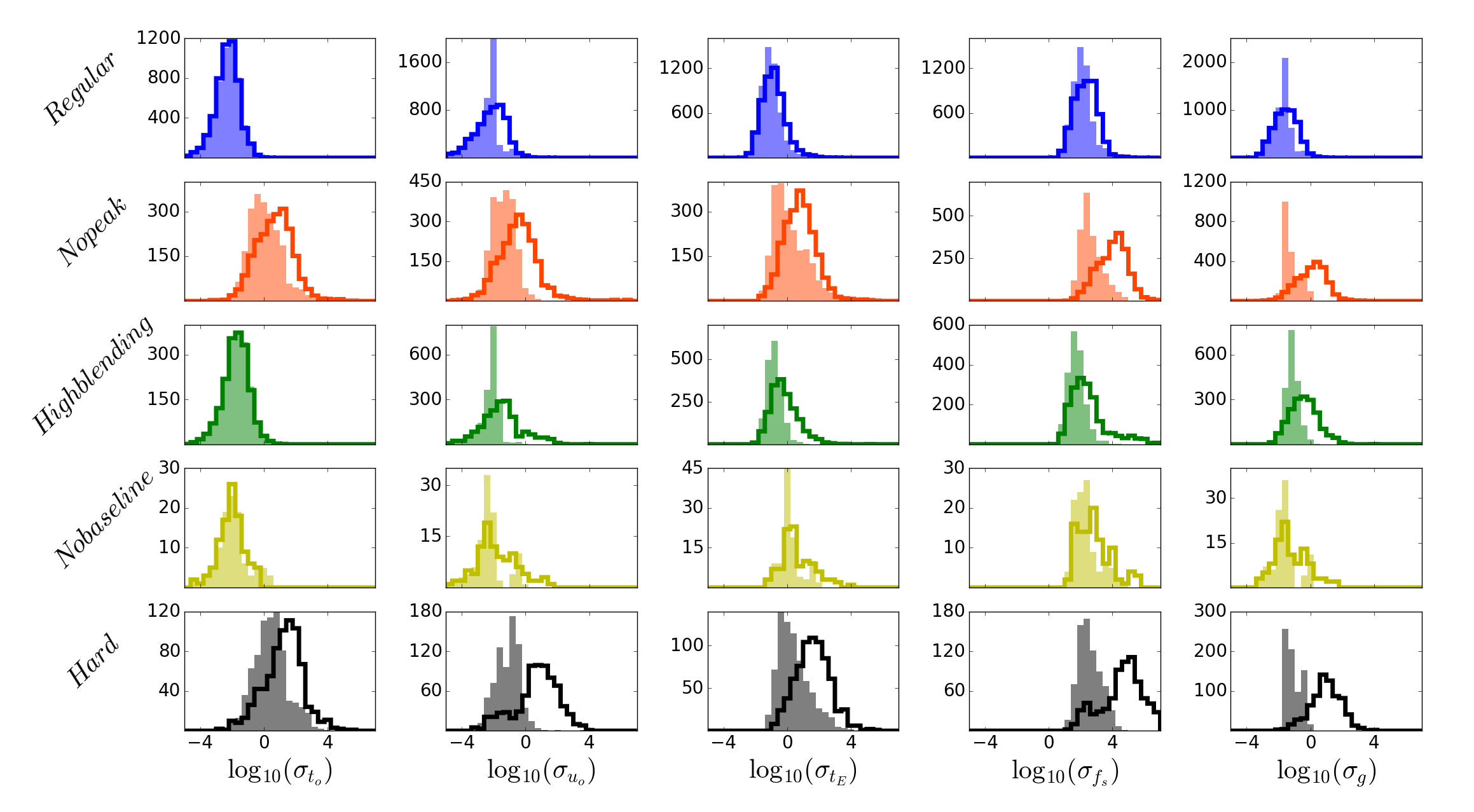}
    \caption{Distributions of parameters errors from pyLIMA 'LM' (thick line histograms) and from pyLIMA 'MCMC' (light plain histograms) for the Space dataset. Events are sorted by categories in line and colors according to Section~\ref{sec:ResultsPSPL}. The two methods estimate similar errobars for well-contrain events, but present severe differences in case of problematic lightcurves. This underlines a data issue rather than a fitting routine problem.}
    \label{fig:errors_hist}
\end{figure*}

Since the 'LM' and 'MCMC' methods estimate parameters uncertainties in a different way, it is informative to compare their respective results on the same datasets. First, we refit all the  Space lightcurves defined in the Section~\ref{sec:Spacesimulations} using the 'MCMC' method. Parameters uncertainties for both methods are visible in the Figure~\ref{fig:errors_hist}. The general trend to notice is that the MCMC and LM methods agree for small uncertainties (i.e well defined events) and disagree for larger uncertainties. The 'LM' method tends to overestimate errors for unconstrained events. This can be explain by the covariance matrix approach of uncertainties estimation, which assume that the $\chi^2$ landscape is parabolic close to a minimum. This hypothesis breaks in case of ill-defined events and therefore a correction factor is needed in the covariance approach. On the other side, uncertainties estimated through the 'MCMC' method look more realistic and stable, regardless of the nature of the event. This of course has a cost : the 'MCMC' fitting time of an event is roughly  2000 times slower than the 'LM' method.

\begin{figure*}    
  \centering
  \includegraphics[width=18cm]{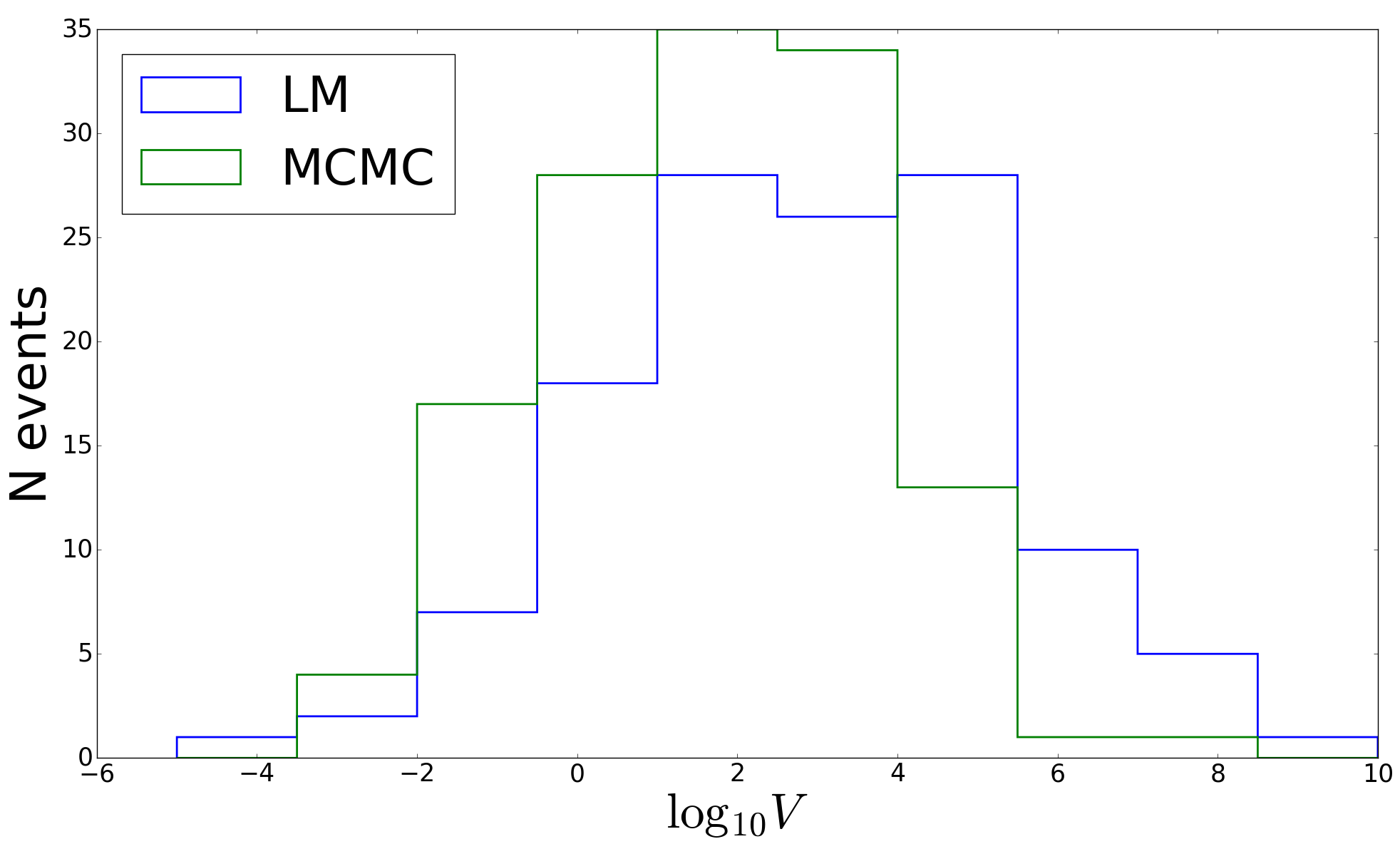}
    \caption{Distributions of error volume from pyLIMA 'LM' (blue) and from pyLIMA 'MCMC' (green) for the OGLE-II dataset. Because this dataset presents microlensing events with strong signals and good photometry, the two methods converge to equivalent fits and errobars. This illustrates the robustness of the 'LM' and 'MCMC' methods on microlensing events without pathologic cases.}
    \label{fig:Vol_total}
\end{figure*}

We decided to conduct a similar study on real data. We use the analysis made on section Section~\ref{sec:realworld} to compare errors coming from the methods 'LM' and 'MCMC' for the OGLE-II fits. However, we decide to change the diagnosis metric and compare the error volume for both datasets \citep{Tsapras2016}:
\begin{equation}
 V = \prod_{i}\sigma_i
\end{equation}
This can be seen on the right of Figure~\ref{fig:Vol_total}. The two distributions present a good agreement (p-valule of the Kolmogorov-Smirnov test higher than 0.9). However, it seems that the error volume estimated form the 'MCMC' looks slightly smaller (the median values for the $\log_{10}V$ are 3.1 and 1.7 for 'LM' and 'MCMC' respectively). This is due to the fact that errors estimated from the 'MCMC' method are not divergent when the events are not that well constrained, whereas the errors from the 'LM' method can grow dramatically (see Section~\ref{sec:simulations}).


\bibliographystyle{mnras}  
\bibliography{biblio_pyLIMA1.bib}

\begin{figure*}    
  \centering
  \includegraphics[width=18cm]{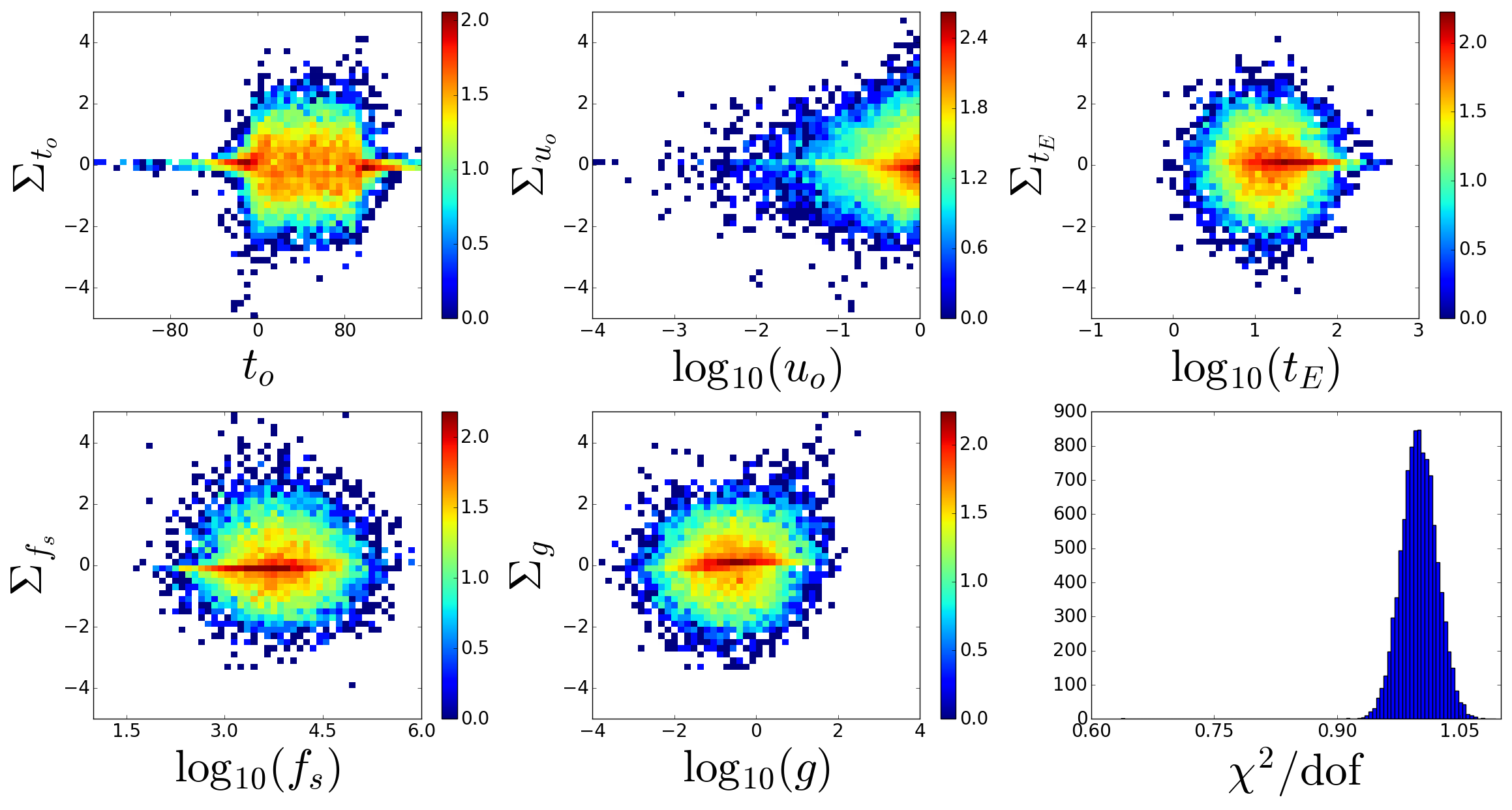}
    \caption{$\Sigma_x$ distributions for the Space dataset. When all parameters present similar results as Figure~\ref{fig:Sigmaground}, the $\chi^2/ \textrm{dof}$ is significantly different. For this dataset, the photomotric precision was not limited to $1\%$, explaining the non-skewed distribution.}
    \label{fig:Sigmaspace}
\end{figure*}
\begin{figure*}   
  \centering
  \includegraphics[width=18cm]{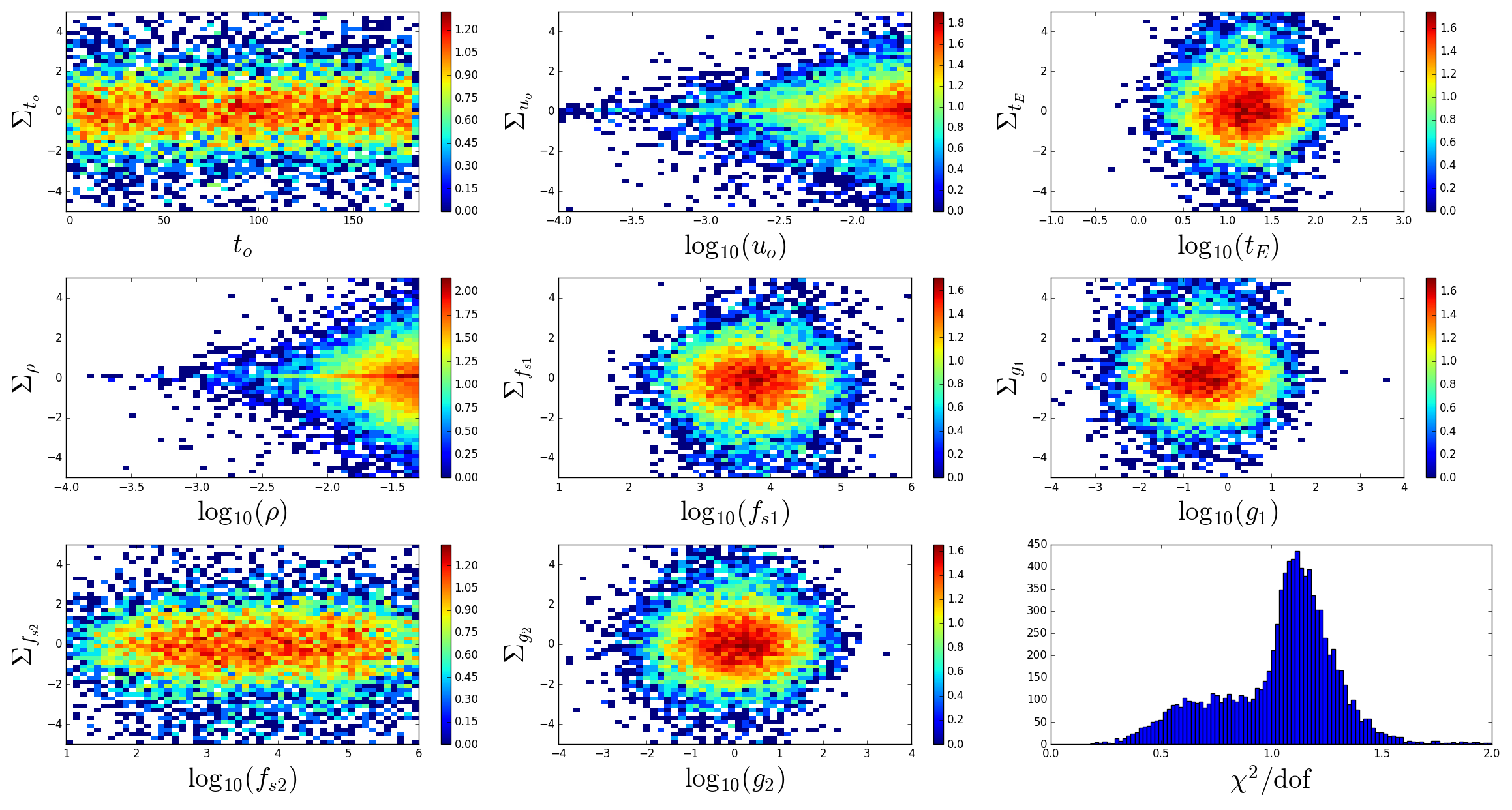}
   \caption{$\Sigma_x$ distributions for the FSPL dataset. In contrary to Ground and Space dataset,  there is no particular trends in the $t_o$ and $u_o$ distributions, because we forced events to peak in the observing windows and because the event peak is well constrained by the Follow-up dataset.}
    \label{fig:Sigmafspl}
\end{figure*}
\begin{figure*}
  \centering
  \includegraphics[width=18cm]{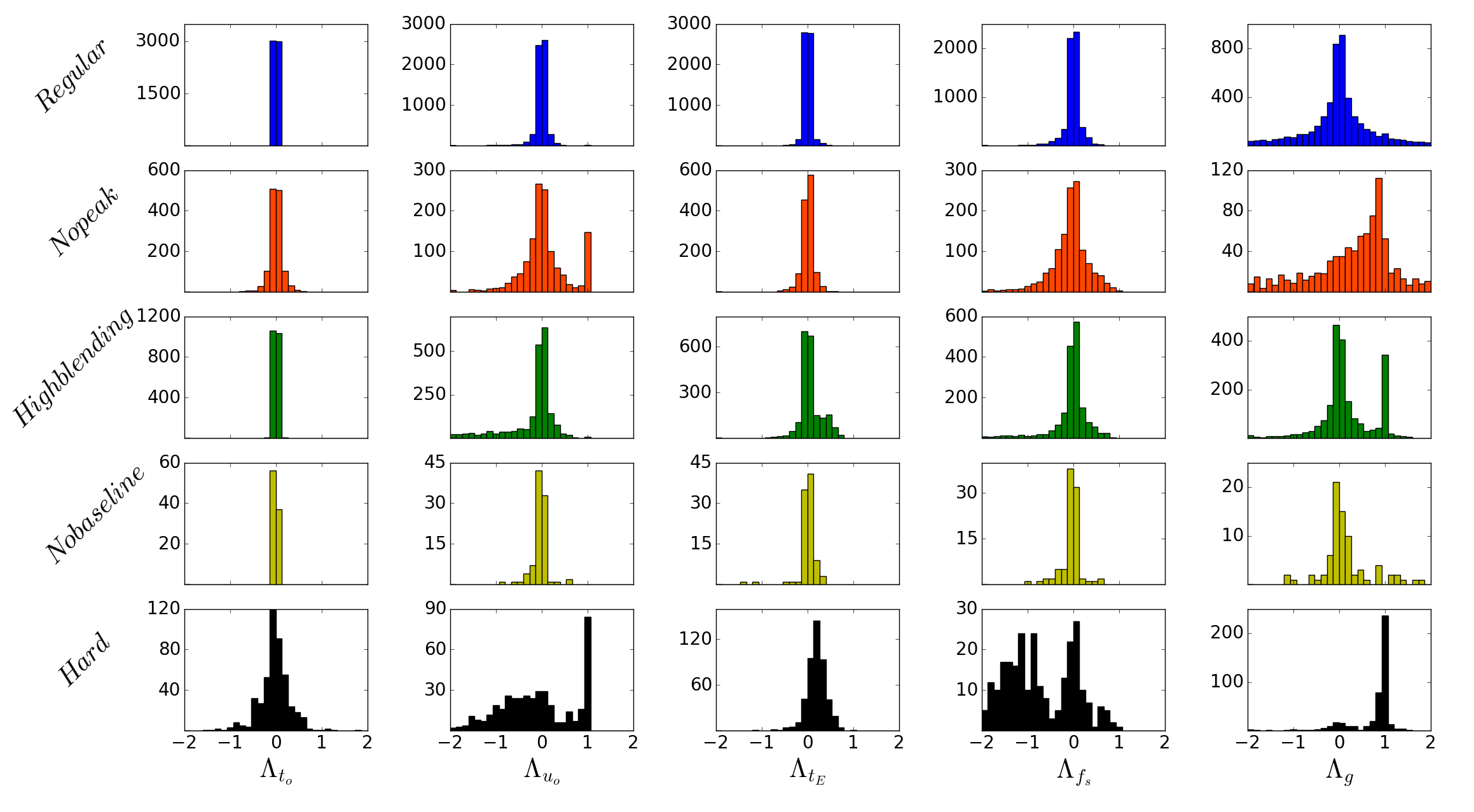}
    \caption{$\Lambda_x$ distributions for the Ground dataset. Again, it is clear that poor fits are due to problematic lightcurves rather than pyLIMA fitting procedure.}
    \label{fig:lambdaground}
\end{figure*}
\begin{figure*}
  \centering
  \includegraphics[width=18cm]{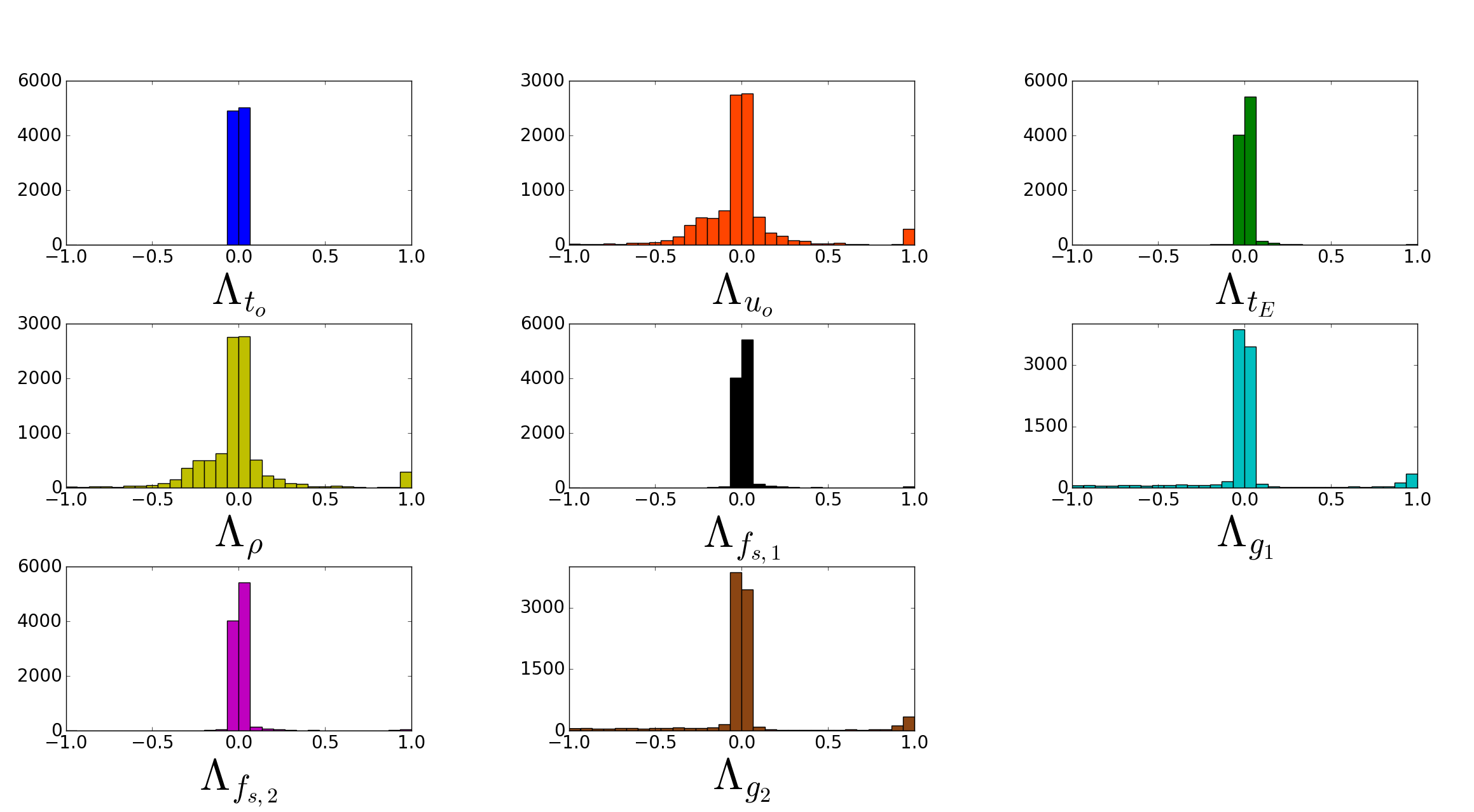}
    \caption{$\Lambda_x$ distributions for the FSPL dataset. Note the excess of events around 1 for the $u_o$, $\rho$, $g_1$ and $g_2$ distributions. These are signatures of fits failures seen in Section~\ref{sec:ResultsFSPL}.}
    \label{fig:lambdafspl}
\end{figure*}

\end{document}